\documentclass[10pt,twocolumn,letterpaper]{article}

\usepackage{cvpr}              
\definecolor{cvprblue}{rgb}{0.21,0.49,0.74}
\usepackage[pagebackref,breaklinks,colorlinks,allcolors=cvprblue]{hyperref}

\def\paperID{*****} 
\def\confName{CVPR}
\def\confYear{2026}

\title{Radiant Foam Rendering on a Graph Processor}

\author{
Zulkhuu Tuya 
\quad Ignacio Alzugaray
\quad Nicholas Fry
\quad Andrew J. Davison
\\
Imperial College London\\
{\tt\small \{zulkhuu.tuya24, i.alzugaray, nicholas.fry20, a.davison\}@imperial.ac.uk}
}

\begin{document}
\maketitle
\begin{abstract}
Many emerging many-core accelerators replace a single large device memory with hundreds to thousands of lightweight cores, each owning only a small local SRAM and exchanging data via explicit on-chip communication. This organization offers high aggregate bandwidth, but it breaks a key assumption behind many volumetric rendering techniques: that rays can randomly access a large, unified scene representation. Rendering efficiently on such hardware therefore requires distributing both data and computation, keeping ray traversal mostly local, and structuring communication into predictable routes.

We present a fully in-SRAM, distributed renderer for the \emph{Radiant Foam} Voronoi-cell volumetric representation on the Graphcore Mk2 IPU (Intelligent Processing Unit), a many-core accelerator with tile-local SRAM and explicit inter-tile communication. Our system shards the scene across tiles and forwards rays between shards through a  hierarchical routing overlay, enabling ray marching entirely from on-chip SRAM with predictable communication.
On Mip-NeRF~360 scenes, the system attains near-interactive throughput (\(\approx\)1\,fps at \mbox{$640\times480$}) with image and depth quality close to the original GPU-based Radiant Foam implementation, while keeping all scene data and ray state in on-chip SRAM.
Beyond demonstrating feasibility, we analyze routing, memory, and scheduling bottlenecks that inform how future distributed-memory accelerators can better support irregular, data-movement-heavy rendering workloads.
\end{abstract}    
\section{Introduction}
\label{sec:intro}

Neural radiance fields (NeRF)~\cite{mildenhall2020nerf} established differentiable volumetric rendering as a route to high-quality novel view synthesis. Since then,  radiance representations have steadily improved reconstruction quality, rendering speed, and ease of use. 

At the same time, applications are moving towards larger, more detailed scenes and interactive use, so methods increasingly trade extra memory for speed. As these models grow, memory footprint and bandwidth, not just raw compute throughput, often become the dominant constraints.

Most current systems rely on modern GPUs and assume a large, unified device memory that rays can access with cheap, irregular random reads. As scenes grow, fitting the representation into GPU memory and sustaining those access patterns becomes increasingly difficult, forcing partitioning or streaming and making memory capacity and bandwidth the main bottlenecks.

\begin{figure}
    \centering
    \includegraphics[width=0.99\linewidth]{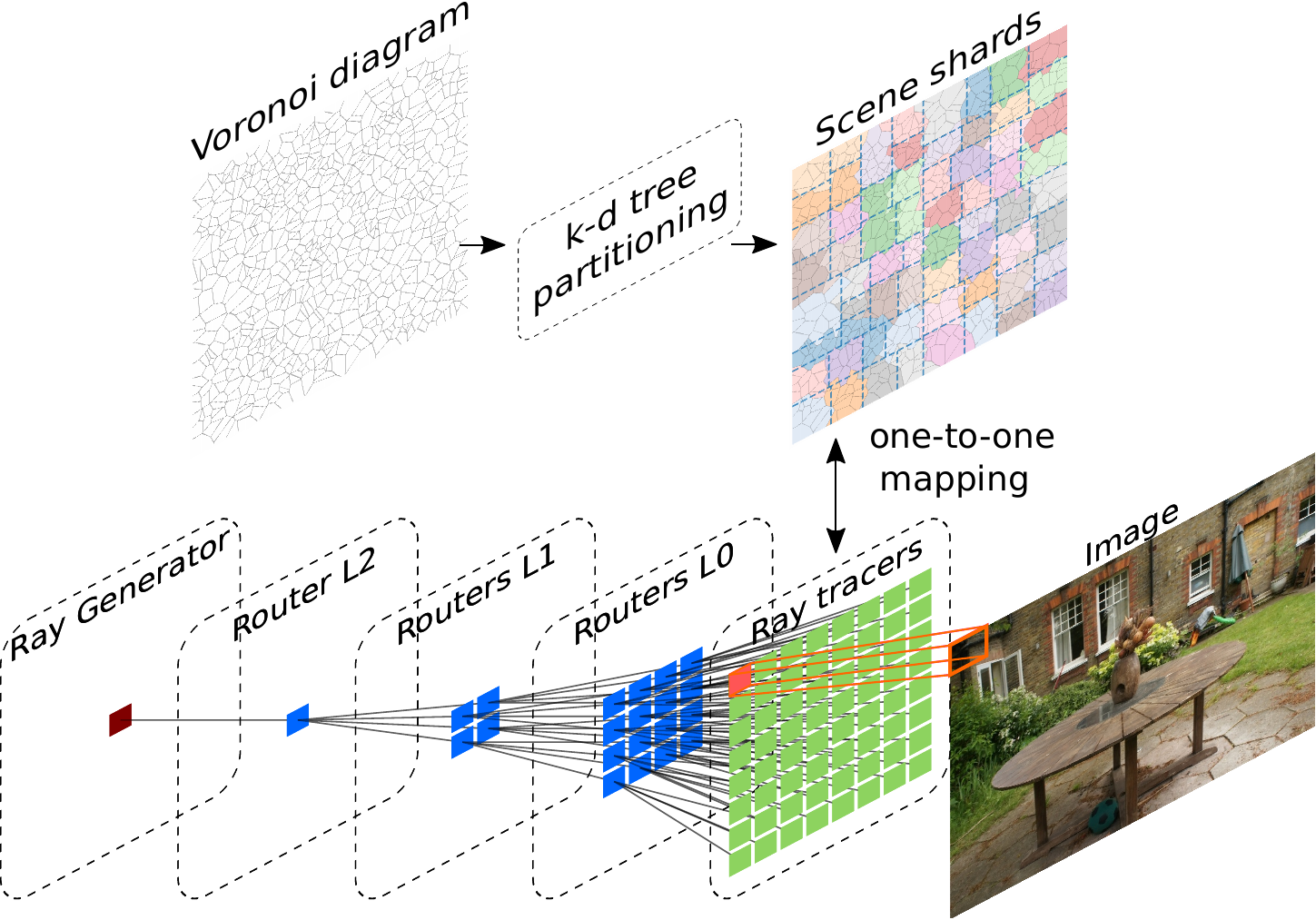}
    \caption[System overview]{System overview. A trained \emph{Radiant Foam} scene is partitioned into SRAM-sized shards and placed on IPU tiles acting as \textcolor[HTML]{8dd35f}{\emph{ray tracers}}. During rendering, rays are forwarded through a sparse \textcolor[HTML]{0066ff}{\emph{quadtree routing overlay}} (illustrated with three router levels for clarity) when they cross shard boundaries, transferring ray state to the tile that owns the next shard. The output image is also partitioned across tiles; each ray’s final pixel value is routed to the tile responsible for that pixel region.}
    \label{fig:teaser}
\end{figure}

In parallel, there is a growing ecosystem of emerging accelerators, such as wafer-scale engines, spatial accelerators, and many-core ML processors~\cite{he2025wafer, soto2023neuralrenderingtpu, pupilli_towards_2023}. Rather than a single large unified memory, these platforms provide many small per-core memories connected by an explicit communication fabric, yielding high aggregate bandwidth at the cost of non-uniform access. Mapping volumetric radiance fields onto such architectures requires rethinking how scene data, rays, and output image are organized and how communication is structured. 
On such architectures, communication is explicit, and performance is best when data placement and data movement can be planned ahead of time rather than data-dependent all-to-all traffic.

We target this class of distributed-memory many-core accelerators: devices in which each core has only a small local SRAM and data exchange is done through explicit message passing or static dataflow, with no large unified device memory. On such hardware, large volumetric radiance fields must be decomposed across tiles and rays must be moved to the cores that own the data they need, since there is no single memory that every tile can access cheaply.

The Graphcore IPU is one concrete instance of this class: a many-core accelerator with over a thousand independently programmable cores, called \emph{tiles}, each with a few hundred kilobytes of tile-local SRAM, connected by an on-chip data exchange fabric rather than a large unified device memory.
While the platform has shown strong performance on selected ML and vision workloads~\cite{kacher2020graphcorec2,finkbeiner2023manycoreipu,gepner2024ipu,Sumeet2022IPUTextDet,Ortiz2020BundleIPU}, existing volumetric rendering methods rely on precisely the features it lacks: global memory and low-cost random access. 
Rendering large volumetric radiance fields on such hardware therefore requires explicit partitioning of the scene and output image across tiles and routing rays through the communication fabric.

We therefore choose a volumetric representation whose ray traversal is local in the representation’s adjacency graph. Specifically, we adopt \emph{Radiant Foam}~\cite{govindarajan_radiant_2025}: because its Voronoi cells form a connected adjacency structure. Rendering a single ray can be viewed as walking the Voronoi diagram: the ray starts in the cell containing the camera position and then advances \emph{cell by cell} by checking only the current cell and its immediate neighbors. 
To execute this ray traversal on an IPU, we partition Voronoi cells into SRAM-sized \emph{shards} and place each shard on a tile (\cref{fig:teaser}).
Whenever a ray moves into a Voronoi cell whose shard is owned by a different tile, its state must be transferred to that tile; similarly, once a ray completes, its pixel contribution needs to be sent to the tile that owns that region of the output image.
Rather than allowing arbitrary all-to-all messaging, we structure these transfers through a static hierarchical routing overlay so that each tile communicates through a small, predictable set of links(\cref{fig:teaser}).

We assume scenes are pre-trained using the original GPU-based Radiant Foam implementation and focus exclusively on rendering on distributed-memory accelerators; enabling training on such hardware is left for future work.

Our contributions are as follows:
\begin{itemize}
  \item A renderer for Radiant Foam tailored to distributed-memory many-core accelerators, implemented on the Graphcore IPU, that keeps both scene data and the output image entirely in tile-local SRAM without requiring global device memory.
  \item A partitioning and routing scheme for Radiant Foam that maps Voronoi cells to SRAM-sized per-tile shards and routes rays between shards through a sparse quadtree overlay, keeping communication predictable.
  \item A compact mixed-precision ray state representation that lowers per-ray SRAM cost, enabling larger buffers for rays per tile.
  \item An empirical evaluation on Mip-NeRF\,360~\cite{barron2022mipnerf360} and Deep Blending~\cite{DeepBlending2018} datasets showing near-interactive rates (around 1\,fps at \(640{\times}480\)) with image and depth quality close to the original GPU-based Radiant Foam implementation, together with an analysis of performance, communication patterns, and load balance on the IPU.
\end{itemize}

\section{Related work}
\label{sec:related}

\subsection{Radiance-field methods}
\label{subsec:rw_radiance_fields}

Neural radiance fields (NeRF) model a continuous 5D function that maps position and view direction to density and radiance, achieving high-fidelity novel views at the cost of many network evaluations per ray~\cite{mildenhall2020nerf}. 
Subsequent works accelerate NeRF by redesigning the representation to better match GPU memory hierarchies and by aggressively pruning unnecessary samples.

3D Gaussian Splatting (3DGS) replaces implicit radiance fields with an explicit set of anisotropic 3D Gaussians and renders them via splatting-based rasterization, enabling real-time performance~\cite{kerbl_3d_2023}. Its explicitness and fast rendering have driven a wave of variants and applications across tasks such as SLAM and robotics, with surveys cataloging the landscape and usage beyond novel-view synthesis~\cite{Matsuki:Murai:etal:CVPR2024,splatam2024,zhu20243dgaussiansplattingrobotics,chen_survey_2024,he2025_gs_survey_2025_apps}.

Motivated by this trend toward explicitness, recent methods directly optimize alternative primitives such as triangles~\cite{Held2025Triangle,held2025trianglesplattingplus}, voxels~\cite{liu2020nsvf,svraster}, and tetrahedra~\cite{gu2024tetsplatting,kulhanek2023tetranerf,lutzow_linprim_2025}.

Despite differing internal mechanics, these approaches share two practical assumptions that shape their engineering: 
(i) a unified device memory where the scene representation and its accelerating structures reside, and 
(ii) a flexible GPU programming model, typically CUDA, with cheap thread-level synchronization and atomics that tolerates irregular but fast random access.
As a result, their throughput gains come from exploiting GPU-friendly data locality and global memory bandwidth, rather than from enforcing strict communication locality between many small, isolated memories as in distributed-SRAM architectures

Complementary to these algorithmic speedups on GPUs, a growing line of work explores custom hardware accelerators for neural rendering. For NeRF, several FPGA- and ASIC-based designs co-design sampling, encoding, and MLP datapaths to reduce latency and power on a single device~\cite{10.1145/3508352.3549380,10.1109/ASP-DAC58780.2024.10473990,10171492,10296239,10.1145/3579371.3589056,10.1145/3695053.3731107}. For 3D Gaussian Splatting, recent processors and architectures specialize sorting, rasterization, and memory hierarchies for Gaussian primitives~\cite{10.1145/3620666.3651385,10848911,10764611,10946749,wang2025accelerating3dgaussiansplatting,pei2025gcc3dgsinferencearchitecture}. These accelerators demonstrate that radiance-field methods benefit from hardware specialization, but they typically entail long design and verification cycles and are tightly coupled to a particular rendering pipeline. Adapting them to new radiance-field architectures or training regimes often requires substantial re-engineering, whereas programmable many-core processors with distributed SRAM, as we explore in this paper with an IPU, can support a wider range of designs via software while still enforcing strict communication locality.

\subsection{Distributed and large-scale rendering}
\label{subsec:rw_distributed_rendering}

Real scenes often exceed a single GPU’s memory, motivating \emph{out-of-core} strategies. In this setting, only the working set is kept resident on the device while the rest of the scene is streamed or paged in from host memory or storage on demand, often using spatial decomposition to limit what must be active per view.
NeRF systems often partition space so that only a small subset of the representation is active per view. \emph{Block-NeRF} trains and serves independent blocks and activates those intersecting the current camera frustum~\cite{tancik2022blocknerf}; \emph{Mega-NeRF} tiles large outdoor environments into regions and schedules training/inference per region to keep the working set bounded~\cite{meganerf2022}. In explicit methods, large-scale 3DGS systems bound the render-time working set by partitioning the scene and using LoD or streaming. CityGaussian~\cite{liu2024citygaussian} and VastGaussian~\cite{lin2024vastgaussian} show this at city- and campus-scale: chunks are paged in as needed, keeping only nearby Gaussians resident. These strategies pair naturally with GPUs’ unified device address space, where blocks can be fetched dynamically with low overhead.

At production scale, distributed renderers follow a related but distinct playbook. Disney’s Hyperion~\cite{burley2018hyperion} handles film-scale assets by sorting and batching rays for coherence while distributing mesh data across CPU nodes with large per-node memory. Recent research systems extend this idea to multi-GPU deployments~\cite{wald2023rayqueue,10.1145/3447807}. \emph{R2E2} achieves low-latency path tracing on terabyte-scale scenes by sharding geometry and textures across thousands of cloud CPUs and forwarding rays to the data~\cite{fouladi_r2e2_2022}. Across these efforts, the target remains triangle-mesh scenes accelerated by BVH(Bounding Volume Hierarchy); throughput at scale is sustained by dynamic communication and ample local memory rather than by strict communication locality or on-chip stores.

\subsection{Emerging AI accelerators beyond GPUs}
\label{subsec:rw_emerging_hardware}

Alongside GPUs and application-specific accelerators, several families of AI accelerators now target large-scale deep learning with different architectures. Wafer-scale engines such as the Cerebras WSE-2 integrate hundreds of thousands of small compute cores and tens of gigabytes of on-chip SRAM on a single wafer, connected by a mesh-like on-chip network~\cite{lie2023cerebras}. Such chips are best viewed as distributed-memory machines, where each core owns a small local store and communication follows explicit routes over the on-wafer fabric~\cite{he2025wafer}. Spatial dataflow accelerators such as Google’s Tensor Processing Units (TPUs) organize computation around systolic matrices attached to high-bandwidth off-chip memory, exposing coarse-grained array-style parallelism rather than the fine-grained thread hierarchy of a GPU~\cite{jouppi2017tpu}. Finally, many-core ML processors such as the Graphcore IPU~\cite{citadel2019ipu,graphcore_ipu_programmers_guide} (see Sec.~\ref{subsec:ipu_background}) populate the chip with over a thousand tiles, each with its own SRAM and statically scheduled communication routes.

What these designs have in common is that performance is obtained by keeping data close to lightweight cores and by constraining communication to predictable patterns over an interconnect fabric, instead of relying on random access into a large, unified device memory as on conventional GPUs. From the perspective of volumetric rendering, they therefore call for representations and algorithms whose working sets fit in tile-local memories and whose communication can be expressed as sparse, bounded exchanges between a small number of neighbors, rather than global look-ups into monolithic acceleration structures.

In this paper, we are inspired by these advances in distributed-memory accelerators and revisit volumetric radiance-field rendering under their constraints. We present an end-to-end renderer on a many-core accelerator with distributed on-chip SRAM, using explicit partitioning and sparse, predictable routing to keep scene data, ray state, and the output image on chip, and we analyze the resulting routing and memory bottlenecks.

\subsection{Rendering on IPUs}
\label{subsec:rw_ipu_rendering}

The Graphcore IPU is a general-purpose many-core accelerator, and graphics and rendering workloads remain comparatively underexplored with only a handful of prior rendering systems have been reported.

Pupilli~\cite{pupilli_towards_2023} presents a screen-space neural path tracer: each tile stores a compact BVH and a replica of the entire scene in its local memory and renders a fixed framebuffer slice. The renderer uses a neural image field (NIF), implemented as a small MLP, to approximate environment lighting on top of simple primitives such as spheres and boxes arranged in Cornell box-style scenes~\cite{cornellbox}. Because all required data fit in each tile’s 624\,kB SRAM, traversal remains local and inter-tile data exchange is avoided. However, replicating the BVH and scene per tile consumes a large fraction of the available SRAM and limits the complexity and scale of scenes that can be handled.

Fry \etal~\cite{fry2024gsplat} port 3D Gaussian Splatting under the same screen-space decomposition: tiles own image regions while a precompiled exchange plan routes projected Gaussians to the correct tiles. The method avoids explicit acceleration structures and benefits from static copies, but can suffer load imbalance when too many Gaussians land in a small image area.

In contrast, we distribute both the scene and the framebuffer across tiles: Radiant Foam’s Voronoi diagram is partitioned spatially over tiles, and the framebuffer is partitioned in screen space, so each tile stores only its local scene shard and image slice. Rays move to whichever tile owns the cell or pixel they need next. This avoids scene data duplication and keeps communication sparse and predictable, even on large Mip-NeRF~360 and Deep Blending scenes.

\section{Preliminaries}
\label{sec:background}

\subsection{Radiant Foam}
\label{subsec:radfoam_background}

Radiant Foam represents the scene as a 3D Voronoi diagram over primal points \(\{p_i\}\).
Given these sites, space is partitioned into convex polyhedral Voronoi cells
\begin{equation}
V_i = \{\,x \in \mathbb{R}^3 \mid \|x - p_i\| \le \|x - p_j\|\ \forall j \neq i \,\}.    
\end{equation}
Each pair of adjacent cells with primal points \(p_i\) and \(p_j\) shares a planar Voronoi face.
Rather than storing these faces explicitly, Radiant Foam recovers them on the fly from the primal points and adjacency information (see \cref{appendix:radfoam_scene_representation} for the full derivation of cell shapes from \(\{p_i\}\) and neighbor indices).
This yields an unstructured adjacency graph: unlike triangle meshes with 3 edges, each Voronoi cell is a convex polyhedron whose number of neighbors is data-dependent.


\begin{figure}
    \centering
    \includegraphics[width=0.75\linewidth]{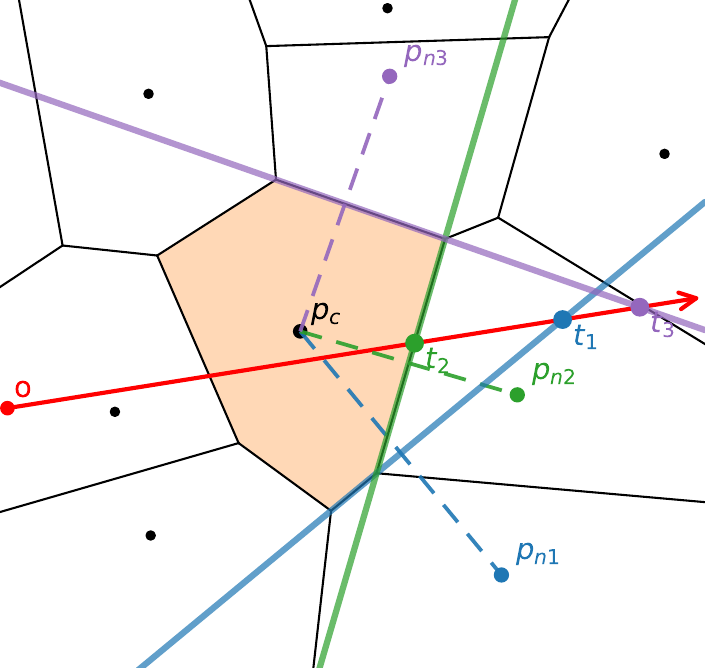}
    \caption[Intersection test in Radiant Foam]{Intersection in Radiant Foam (2D illustration). Primal points are shown as dots and labelled \(p_c\) for the current cell and \(p_{n1}, p_{n2}, p_{n3}\) for neighboring cells. Their Voronoi cells appear as polygonal regions separated by straight line segments. A single ray (red arrow) traverses multiple neighboring cells: within the current cell, it intersects several candidate faces, with the corresponding distances \(t_1, t_2, t_3\) calculated. The ray then crosses the face with the smallest positive distance (here \(t_2\)) into the adjacent cell associated with \(p_{n2}\).}
    \label{fig:intersection_test_radfoam}
\end{figure}

During rendering, a ray determines its next cell by intersecting the faces of the current cell, choosing the smallest positive crossing distance (\cref{fig:intersection_test_radfoam}), and stepping through that face into the corresponding neighboring cell.
Thus, the ray’s path is a sequence of neighboring cells, and ray traversal is essentially a local, topological walk on the adjacency graph.
Along this path, the ray accumulates color and transmittance segment by segment.
For Radiant Foam's Voronoi cells with constant density and colors, the volume-rendering integral along a ray can be written as a sum over the \(N\) segments with constant field values:
\begin{equation}
c_r = \sum_{i=1}^{N} T_{i-1}\,\bigl(1 - e^{-\sigma_i \delta_i}\bigr)\, c_i,
\label{eq:rf_compositing_color}
\end{equation}
where \(\delta_i\) is the length of segment \(i\) and
\begin{equation}
T_i = \prod_{j=1}^{i} e^{-\sigma_j \delta_j}, \qquad T_0 = 1
\label{eq:rf_compositing_transmittance}
\end{equation}
is the accumulated transmittance up to the start of next segment \(i+1\).

Ray's traversal through the Voronoi diagram depends only on local cell adjacency without requiring a global acceleration structure such as a BVH.
The original Radiant Foam implementation targets a single GPU with a large, unified device memory, where all Voronoi cells and their adjacency can be stored in one address space.
Scaling this representation to larger scenes or to accelerators without shared device memory requires explicitly partitioning the Voronoi cells and routing rays between partitions.
In the next subsection, we describe the Graphcore IPU as a representative distributed-SRAM architecture on which we build such a design.

\subsection{Graphcore IPU}
\label{subsec:ipu_background}

In this paper, we use the Graphcore Mk2 IPU, a graph processor with 1472 small processor ``tiles''.
Each tile is a simple core with 624\,kB of its own local SRAM and no shared cache or global device memory: a tile can read and write only its own SRAM.
To exchange data, tiles send messages over a high-bandwidth on-chip network rather than issuing loads and stores into a single large address space, as on a GPU.

Programs are expressed in Graphcore’s Poplar framework as kernels mapped to tiles and operating on explicitly placed tensors~\cite{graphcore_ipu_programmers_guide,citadel2019ipu}. 
Execution proceeds in bulk-synchronous phases: in each phase, tiles (i) compute on their local data, (ii) reach a global synchronization point, and then (iii) exchange the next slice of data along pre-defined routes.
Communication between tiles is scheduled at compile time, so exchange routes and buffer sizes are fixed ahead of execution, requiring bounded communication degrees and static message patterns.
Within a tile, each kernel is executed by a small group of worker threads that share tile-local SRAM. So, concurrent updates must be organized so that each worker writes to disjoint memory regions between synchronization points.

IPU performance therefore depends on several factors: how the scene and computation are mapped onto tiles so that each tile fits within its 624\,kB SRAM and communication remains limited and how exchange buffers are sized for steady throughput~\cite{graphcore_ipu_programmers_guide}. 
In the volumetric rendering setting we study in this paper, both data and computation must live in memory that is distributed per tile.
Scene partitions, ray buffers, and per-pixel outputs (color and depth) all reside in tile-local SRAM; 
there is no single monolithic scene acceleration structure stored in a unified memory that all rays can index arbitrarily.
Ray traversal must therefore remain local so that cross-tile communication stays sparse and statically schedulable, with a limited number of connections per tile.


These constraints align with Radiant Foam’s local, face-to-face adjacency and motivate the fully in-SRAM, distributed renderer we describe next.

\section{Method}
\label{sec:method}

\subsection{System Overview}
Our goal is to render Radiant Foam scenes entirely from tile-local SRAM on Graphcore IPUs, without relying on external DRAM.
This requires expressing both the scene and the output image in terms of per-tile data and computation, distributing them over many tiles, and replacing irregular neighbor-to-neighbor communication with a structured routing scheme.

We organize the IPU tiles into three roles (\cref{fig:teaser}). \emph{Ray tracer} tiles each store a shard of the Radiant Foam scene and a corresponding slice of the framebuffer, and perform ray marching on rays traversing their local Voronoi cells. \emph{Router} tiles forward rays to their next destination. A single \emph{ray generator} tile creates new rays from the current camera parameters. All tiles execute in parallel on the IPU: tracers advance rays through local data, routers forward rays toward their next destination, and the generator injects new rays. Periodically, the host reads back the distributed framebuffer slices and assembles the final image.


\subsection{Scene representation and partitioning}
\label{sec:scene_partitioning}
A trained Radiant Foam scene consists of a set of Voronoi cells, each carrying a primal point and its neighbor indices, density, and view-dependent color parameterized by spherical harmonics (SH). Directly storing all 45 SH coefficients per cell exceeds the 624\,kB SRAM budget on a single tile. We therefore discard higher-order SHs and retain only constant RGB coefficients, trading some view-dependence for a more compact representation that can be distributed over tiles.

\begin{figure}
    \centering
    \includegraphics[width=0.95\linewidth]{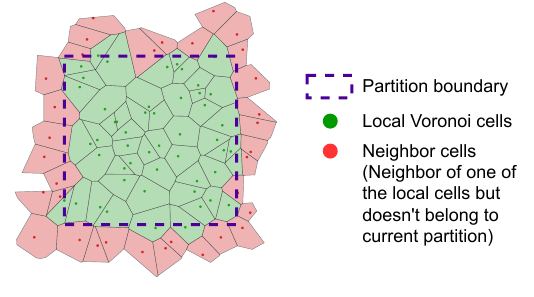}
    \caption[Partition of the Voronoi diagram]{Partition of the Voronoi diagram. \textcolor{green}{Local cells} in a partition require access to the positions of \textcolor{red}{neighbor cells} across boundaries in order to reconstruct their Voronoi cell shapes during rendering.}
    \label{fig:local_neighborhood_filled}
\end{figure}

To map this representation onto the IPU, we first partition the primal points with a balanced k-d tree of depth 10, yielding $1024$ scene shards. Each shard becomes a \emph{scene partition} that will be assigned to one tile. Because Voronoi faces are defined implicitly by neighboring sites, cells near partition boundaries require information about neighbors that may live in adjacent partitions. For each partition, we store below as shown in~\cref{fig:local_neighborhood_filled}:
\begin{itemize}
  \item \textbf{Local cells:} primal points and their attributes that belong to this partition;
  \item \textbf{Neighbor cells:} positions (and identifiers) of cells that lie just across the partition boundary;
  \item \textbf{Adjacency lists:} compact lists of neighbor indices that allow reconstruction of local Voronoi faces (see \cref{appendix:radfoam_scene_representation} for full derivation).
\end{itemize}

During rendering, each tile reconstructs the shape of its local cells using only its local cell and neighbor lists. When a ray exits a local cell and crosses a partition boundary, the tracer uses a precomputed ``next partition ID'' to hand the ray off to the appropriate neighbor partition.


\subsection{Routing layer and ray routers}
\label{subsec:routing_layer}
Radiant Foam’s traversal cost per step depends only on the number of faces of the current cell rather than on the size of the full scene, so it is natural to replace a global acceleration structure with a hierarchy of small, local routing decisions.
A naive implementation would allow each partition to send rays directly to any neighboring partition.
On the IPU this is undesirable: leaf-to-leaf communication patterns would be irregular, buffers would need to be allocated for many possible neighbors, and Poplar expects communication patterns to be statically declared.

\begin{figure}
    \centering
    \includegraphics[width=0.99\linewidth]{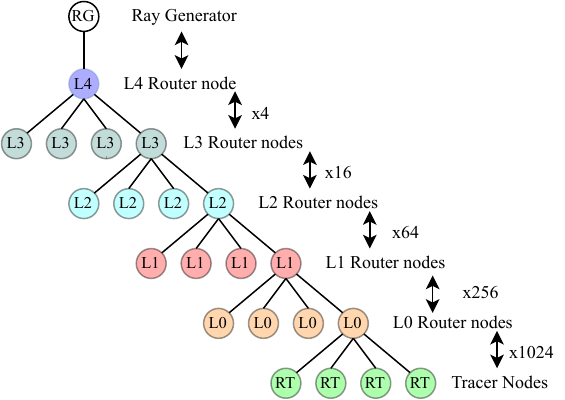}
    \caption[Quadtree-based routing overlay]{Quadtree-based routing overlay(routing tree). Only rightmost branches are shown for clarity.}
    \label{fig:render_tree}
\end{figure}

Instead, we introduce a \emph{routing layer} built from dedicated tiles arranged as a quadtree (\cref{fig:render_tree}). The leaves of this tree are the $1024$ ray tracers; internal nodes are \emph{ray routers}. We measure routing distance in \emph{hops}, where one hop is a transfer of a ray’s payload across a single edge of the routing tree (between two directly connected tiles: tracer$\leftrightarrow$router or router$\leftrightarrow$router) during an exchange. 
With $L{=}5$ levels, the maximum root-to-leaf depth is 5, so the worst-case leaf-to-leaf path is
\(
d_{\max} = 2L = 10
\)
hops (up to five hops up and five down).

This quadtree overlay has several advantages. First, each router has fixed degree (four children and one parent), so its SRAM budget is split across only five links, allowing substantially larger per-link buffers than an all-to-all design. Second, the topology is regular and known at compile time, aligning with Poplar’s preference for statically declared communication graphs. Third, because we partition the scene with a balanced k-d tree, spatially nearby shards tend to fall under the same quadtree branches, which reduces the average hop count and communication distance. Finally, the Mk2 tile budget (1472 tiles) comfortably accommodates the routing overlay and tracing roles: we allocate 1024 tiles as ray tracers and 341 tiles as routers, leaving a small remainder for ray generation and I/O.

Tree overlays concentrate traffic at the upper levels, which would ideally be widened (Fat-Tree~\cite{leiserson1985fat_transactions}) to provide more bandwidth near the root. Because IPU tiles are homogeneous with fixed SRAM memory for data exchange, widening would require replicating upper-level routers across multiple tiles and distributing traffic across them.
This makes congestion at higher levels a potential bottleneck; we analyze the resulting hop statistics in \cref{sec:evaluation}.

\paragraph{Two-pass routing on IPU.}
Each ray router tile receives rays from its parent and four children and must forward them to the correct output lane based on the next destination partition. On the IPU, multiple worker threads operate in parallel and output buffers have fixed capacity. Since Poplar does not provide locks or atomic operations, naive concurrent writes would cause data races and overflows.

\begin{figure}
  \centering
  \includegraphics[width=0.995\linewidth]{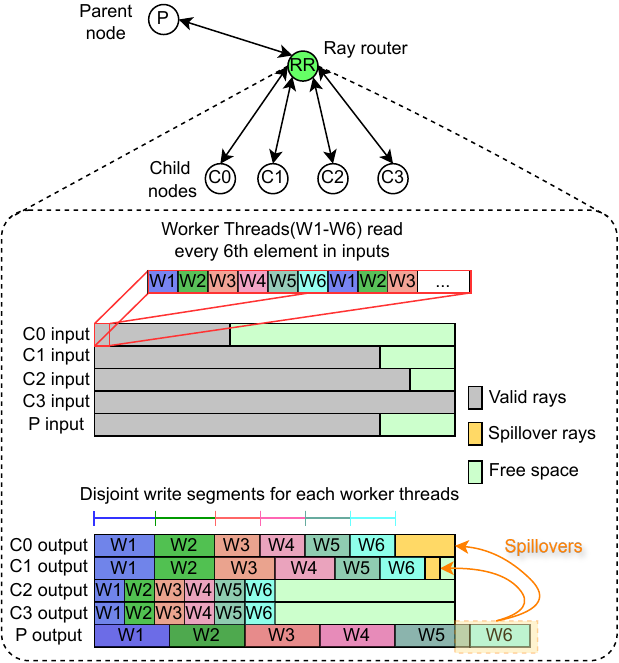}
  \caption[Operation of a ray router tile]{Operation of a ray router tile. Rays are received from the parent and four children and forwarded to the correct output lane based on their target partition.}
  \label{fig:ray_router}
\end{figure}

We therefore implement routing as a two-pass, barrier-synchronized forwarder (\cref{fig:ray_router}):
\begin{enumerate}
  \item \textbf{Counting pass.} Each worker scans its assigned portion of the input buffer and counts how many rays will be sent to each output lane. These per-worker counts are aggregated to compute disjoint write ranges for all workers and all lanes.
  \item \textbf{Forwarding pass.} Workers copy rays into their pre-assigned segments of the output buffers. Since the ranges are disjoint, no two workers write to the same memory region.
\end{enumerate}

To keep active rays packed at the head of each buffer, routers compact valid entries and invalidate the tail using a special marker in the destination field. This allows both routers and tracers to detect empty ranges quickly and terminate early.

\subsection{Ray representation}
\label{subsec:ray_representation}
A ray must carry all information needed for its traversal and color compositing: at minimum, distance $t$, transmittance $T$, accumulated color $(r,g,b)$, and \emph{destination ID} consisting of the target shard ID and the entry cell within that shard. In a naive layout, we might also store ray's origin and direction explicitly as three-component 32-bit floats each. This quickly becomes too large for holding many rays in a SRAM buffer.

On the IPU we can reconstruct origin and direction from the camera parameters and pixel index, which are constant over a frame. We therefore store only fields that cannot be reconstructed from camera state and the pixel index $(x,y)$. The camera origin and per-pixel direction are derived on device from the broadcast view/projection matrices and pixel coordinate $(x,y)$, so they are not carried in the payload.

\begin{table}
\centering
\caption{Per-ray payload variants used in our renderer.}
\label{tab:ray_variants}
\begin{tabular}{lccc}
\toprule
Layout & $t,T$ precision & RGB precision & Size \\
\midrule
Default         & fp32 & fp32 & 28\,B \\
Mixed           & fp16 & fp32 & 24\,B \\
All-half+depth  & fp16 & fp16 & 20\,B \\
\bottomrule
\end{tabular}
\end{table}

We consider several payload layouts summarized in~\cref{tab:ray_variants}.
In the \emph{Default} layout, the current distance along the ray $t$, the transmittance $T$, and the accumulated color $(r,g,b)$ are stored as standard 32-bit (\texttt{float}), together with 16-bit integer fields for destination shard ID and local-cell indices. The \emph{Mixed} layout stores $(t,T)$ using Poplar’s built-in 16-bit floating-point type \texttt{half} (fp16), while keeping RGB in 32-bit floats, to increase buffer capacity. The final configuration, \emph{All-half+depth}, stores all floating-point fields in \texttt{half} and pads the payload to meet the IPU’s 4-byte exchange granularity. We fill the resulting 2\,B pad with a 16-bit depth value (estimated at 50\% transmittance quantile), yielding a 20\,B, 4-byte–aligned payload that also supports lightweight depth visualization.

\subsection{Ray tracers}
\begin{figure}
    \centering
    \includegraphics[width=0.92\linewidth]{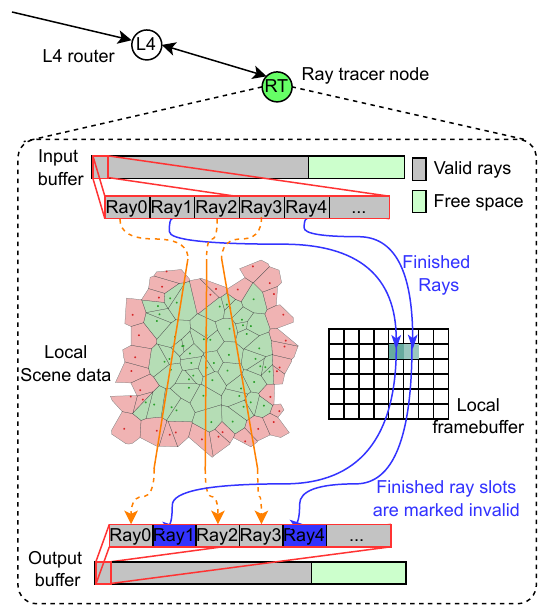}
    \caption[Operation of a ray tracer tile]{Operation of a ray tracer tile. Rays are marched through local Voronoi cells and forwarded either back into the routing tree or to the local framebuffer slice.}
    \label{fig:raytracer}
\end{figure}

Each leaf tile in the \emph{routing overlay} acts as a ray tracer: it stores one scene shard (local and neighbor Voronoi cells) and a local framebuffer slice, and advances rays through its Voronoi cells (\cref{fig:raytracer}).

During execution, six worker threads process disjoint index ranges of the input buffer. For each active ray, a worker:
\begin{enumerate}
  \item Reconstructs the ray origin and direction from the camera parameters and pixel coordinates $(x,y)$.
  \item Uses the current local cell index to look up density and color for that Voronoi cell.
  \item Marches the ray through the cell, updating transmittance $T$ and accumulated color $(r,g,b)$.
  \item Determines whether the ray terminates (e.g.\ $T$ below a threshold or leaving the scene bounds), crosses into a neighboring cell within the same shard, or needs to cross a partition boundary.
\end{enumerate}
If the ray remains within the shard, the tracer updates its local cell index and continues processing it on the same tile. 
If the ray crosses a shard boundary, the tracer updates its destination ID to match it and enqueues the ray into an outgoing buffer; during the next communication exchange, this buffer is copied to the routing overlay and the ray is forwarded to the tile that owns the destination shard. When a ray terminates, its final pixel contribution is accumulated into the tile-local framebuffer slice.

To remain race-free without atomics, each worker writes results back to the same indices it consumed, so all reads and writes are purely index-based. Finished rays are marked as invalid in the output buffer after their contributions are accumulated, allowing indices to remain consistent across workers and enabling deterministic execution.

\subsection{Ray generator and batched injection}
\label{subsec:ray_generator}
\begin{figure}
    \centering
    \includegraphics[width=0.995\linewidth]{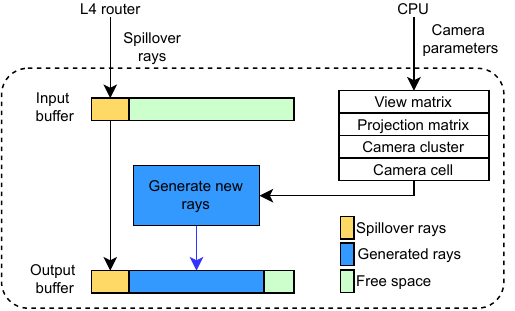}
    \caption[Ray generator]{Ray generator: new rays are initialized from camera state and injected at the root of the routing tree. The generator also acts as a spill buffer for the top-level router.}
    \label{fig:ray_generator}
\end{figure}

A single generator tile at the quadtree root creates new rays from the current camera parameters and injects them into the routing overlay (\cref{fig:ray_generator}). At the start of a batch, the host broadcasts the camera state (view/projection matrices, intrinsics, and the current camera cell) to all tiles. The generator then initializes rays and forwards them into the top level router tile.

The routing overlay, however, cannot absorb a full frame in a single pass without saturating tile buffers at upper levels. We therefore inject rays in structured batches, using row- or column-wise scans of the image plane. Between batches, the system can insert short \emph{drain windows} during which the generator emits no new rays, allowing currently in-flight rays to finish and avoid congestion in routers. The generator also serves as an overflow target for the level-4 router(L4 in~\cref{fig:render_tree}): when a top-level router becomes saturated, excess rays are temporarily spilled into the generator’s input queue, improving robustness under bursty workloads.

\begin{figure*}
  \centering
  \begin{subfigure}[t]{0.495\textwidth}
    \includegraphics[width=\linewidth]{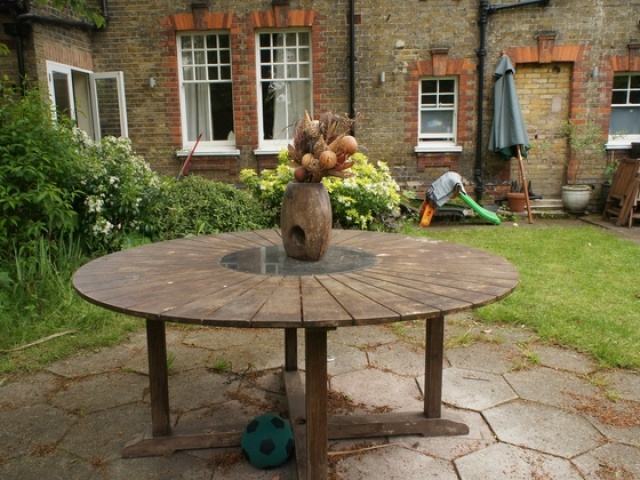}
    \caption{Original Radiant Foam rendering results on GPU)}
  \end{subfigure}\hfill
  \begin{subfigure}[t]{0.495\textwidth}
    \includegraphics[width=\linewidth]{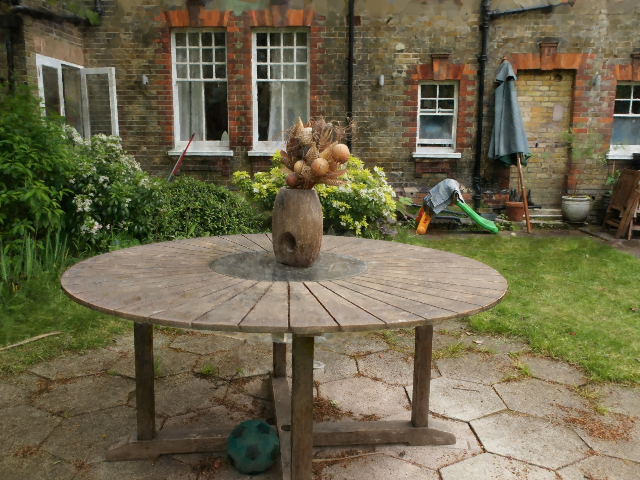}
    \caption{Ours (IPU, in-SRAM)}
  \end{subfigure}
  \caption{RGB reconstruction on the Mip-NeRF360 \emph{Garden} scene. Our in-SRAM IPU renderer closely matches the original Radiant Foam implementation while keeping the entire model and all ray state on chip.}
  \label{fig:rgb_quality_min}
\end{figure*}

\subsection{Distributed framebuffer}
The contributions of finished rays must be accumulated into an image that is orders of magnitude larger than a single tile’s SRAM. For example, a $640{\times}480$ RGB framebuffer requires roughly 900\,kB, exceeding the 624\,kB memory of a tile even before accounting for scene data and buffers. The framebuffer must therefore be distributed.

Ray tracers are a natural place to store framebuffer slices: they are the most numerous tiles, they are where most rays terminate, and they retain enough memory after storing their scene shards. We evaluated two strategies:
\begin{itemize}
  \item \textbf{Local storage.} Each tile keeps finished rays in a dedicated buffer which is read back directly by the CPU. This has minimal on-device processing overhead but requires large per-tile buffers and frequent CPU readbacks to avoid overflow.
  \item \textbf{Screen-space partitioning.} The framebuffer is divided into fixed screen-space slices, similar to the screen-partitioning scheme used in previous systems. When a ray finishes, it is routed through the quadtree to the tile responsible for its pixel slice. Each tile maintains a buffer exactly sized to its framebuffer region, eliminating overflow risk. The CPU can read back the distributed framebuffer once per frame or once every few iterations.
\end{itemize}

Experiments showed that screen-space partitioning is significantly faster: the additional routing traffic from finished rays was negligible compared to the cost of frequent large readbacks required by the local-storage scheme. In our renderer we therefore adopt the screen-space partitioning strategy and let ray tracers serve both as scene partitions and framebuffer slice holders.

\subsection{Execution on IPU}
\label{subscn:execution_on_ipu}
The IPU follows a bulk-synchronous parallel (BSP) model: computation proceeds in phases separated by global synchronization points where data exchange occurs. We exploit this by grouping all tiles (ray tracers, routers, and generator) into a single concurrent compute step that execute parallel, then alternating compute and exchange programs. The resulting IPU program consists of the following subprograms or kernels:
\begin{enumerate}
  \item \textbf{Camera broadcast.} The host sends updated camera parameters to the generator and broadcasts them to all tiles.
  \item \textbf{Compute step.} Ray tracers advance rays through local cells and accumulate finished rays into their framebuffer slices; routers perform two-pass forwarding; the generator emits new primary rays for the current batch.
  \item \textbf{Data exchange.} Rays are copied from output buffers to the corresponding input buffers of their parent or child nodes in the routing tree. Many small copies are merged into a single Poplar exchange program, which the compiler schedules efficiently across the IPU-Exchange fabric.
  \item \textbf{Repeat.} Steps 2 and 3 are repeated $RC$ times per frame, where $RC$ (repeat count) controls how many compute–exchange iterations are performed before reading back the framebuffer.
  \item \textbf{Framebuffer readback.} After $RC$ iterations, the host reads back the distributed framebuffer slices from all ray tracers and assembles the final image.
\end{enumerate}

Choosing $RC$ trades off host–device overhead against latency: too few iterations underutilize the fabric due to frequent host calls, while too many yield diminishing returns because buffers eventually become saturated. We study this trade-off empirically in \cref{sec:evaluation}.

\begin{figure*}
\centering
\begin{subfigure}[t]{0.33\textwidth}
    \centering
    \includegraphics[width=\textwidth]{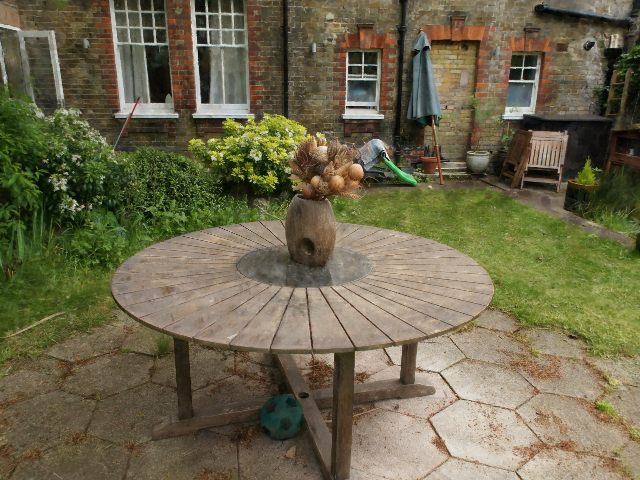}
    \caption{Full precision}
\end{subfigure}\hfill
\begin{subfigure}[t]{0.33\textwidth}
    \centering
    \includegraphics[width=\textwidth]{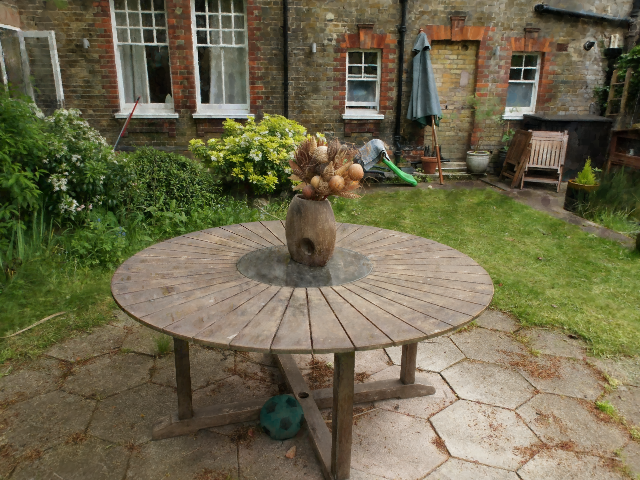}
    \caption{Mixed precision}
\end{subfigure}\hfill
\begin{subfigure}[t]{0.33\textwidth}
    \centering
    \includegraphics[width=\textwidth]{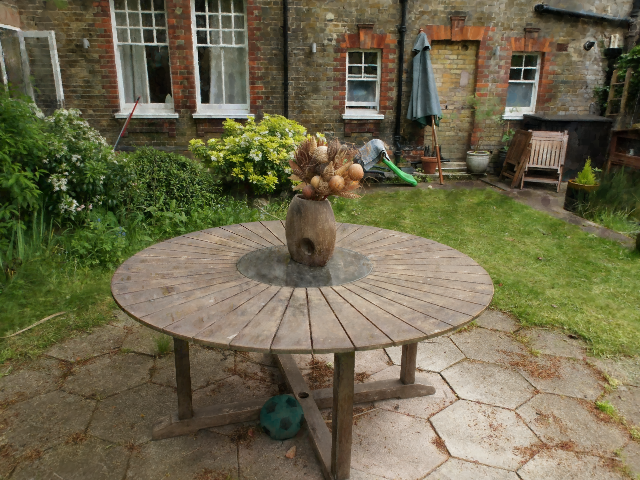}
    \caption{Half precision}
\end{subfigure}
\caption{\emph{Garden} at different payload precisions. Visual differences are imperceptible, matching the PSNR/SSIM in ~\cref{tab:psnr_ssim_appendix}.}
\label{fig:precision_result_compare_appendix}
\end{figure*}

\section{Evaluation}
\label{sec:evaluation}

We evaluate our fully in-SRAM, distributed Radiant Foam renderer with three questions in mind:
(1) how closely it matches the RGB and depth quality of the original Radiant Foam implementation,
(2) whether full Mip-NeRF360 scenes fit comfortably within tile-local SRAM, and
(3) what end-to-end throughput and routing behavior it achieves under realistic batching and precision choices.
Unless noted otherwise, all timings are measured at \(640{\times}480\) resolution on Mip-NeRF360 scenes, using the IPU configuration described in~\cref{subsec:ipu_background} and ~\cref{sec:method}.

\subsection{Reconstruction quality}
\label{subsec:rendering_quality}

\paragraph{Visual quality.}
\cref{fig:rgb_quality_min} compares our IPU implementation to the Radiant Foam ground truth on the \emph{Garden} scene.
Despite running entirely from on-chip SRAM with simplified per-cell color (constant RGB instead of spherical harmonics), the IPU renderer preserves fine detail with only minor view-dependent effects differences.

\paragraph{Image fidelity vs. ray-payload precision.}
To quantify reconstruction quality, we compare our in-SRAM renderer against a full-precision Radiant Foam implementation on several Mip-NeRF360 views while varying the ray-payload precision (~\cref{subsec:ray_representation}).
\cref{tab:psnr_ssim_appendix} reports PSNR and SSIM for the \emph{Garden} scene.
Even with the compact half-precision payload used in our final configuration, PSNR remains above 55\,dB and SSIM is effectively 1.0, and visual differences are imperceptible.

\begin{table}
\centering
\caption{PSNR/SSIM vs full-precision output for the \emph{Garden} scene.}
\label{tab:psnr_ssim_appendix}
\begin{tabular}{|l|c|c|}
\hline
Comparison & PSNR (dB) & SSIM \\
\hline
Full vs Mixed & 56.09 & 1.000 \\
Full vs Half  & 55.91 & 1.000 \\
\hline
\end{tabular}
\end{table}

\cref{fig:precision_result_compare_appendix} visualizes the same comparison: full-, mixed-, and half-precision payloads produce outputs that are visually indistinguishable.

\begin{figure*}
  \centering
  \begin{subfigure}[t]{0.49\textwidth}
    \includegraphics[width=\linewidth]{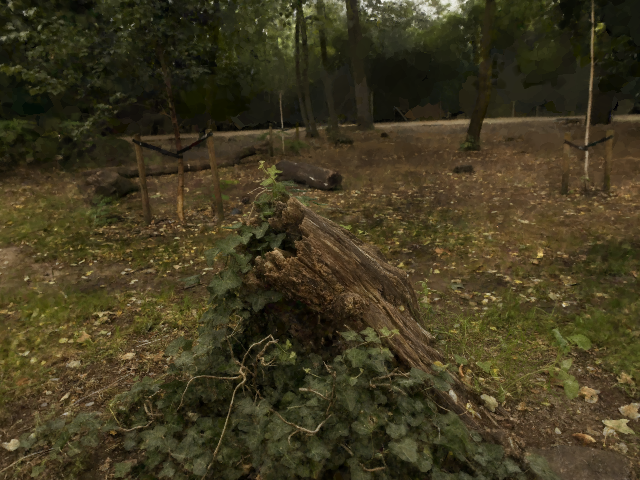}
    \caption{RGB}
  \end{subfigure}\hfill
  \begin{subfigure}[t]{0.49\textwidth}
    \includegraphics[width=\linewidth]{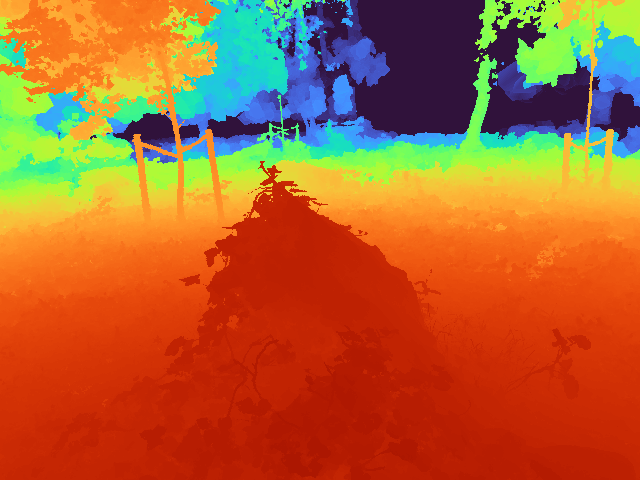}
    \caption{Depth (50\% quantile)}
  \end{subfigure}
  \caption{Color and depth on the \emph{Stump} scene. Depth is computed at the 50\% transmittance quantile and tracks scene geometry smoothly at a resolution of 640$\times$480 pixels.}
  \label{fig:rgb_depth_single}
\end{figure*}

\paragraph{Depth maps.}
The same ray-marched transmittance used for color also produces a depth estimate at a fixed transmittance quantile.
\cref{fig:rgb_depth_single} shows RGB and 50\% transmittance-quantile depth on the \emph{Stump} scene.
Depth aligns well with geometry and is stable across viewpoints, indicating that the Voronoi-cell traversal and accumulation behave as intended.
Additional paired RGB/depth results on other scenes are provided in~\cref{appendix:additional_qualitative}.

\subsection{Scene partitioning and on-chip memory footprint}
\label{subsec:partition_eval}

We next verify that whether large Mip-NeRF360 Radiant Foam scenes can be partitioned to respect the Mk2 IPU’s 624\,kB per-tile SRAM limit while preserving spatial locality.

\begin{figure}
    \centering
    \includegraphics[width=\linewidth]{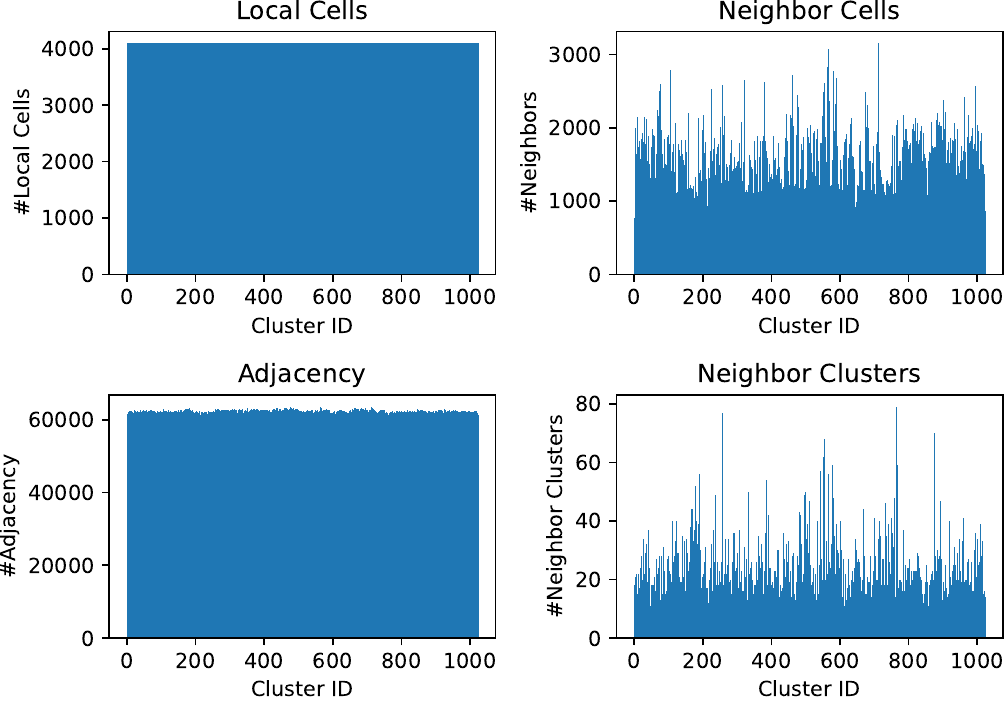}
    \caption{K-d tree partitioning of \emph{Bicycle} scene into 1024 clusters (scene partitions). Each cluster corresponds to one leaf of the k-d tree and becomes a scene partition on a ray-tracer tile. For each cluster, the four subplots report the number of local cells ($N_\text{local}$), neighbor cells ($N_\text{neighbor}$), adjacency entries ($N_\text{adjacency}$), and neighboring clusters, respectively.}
    \label{fig:bicycle_kdtree_partition_result}
\end{figure}

Using a 10-level k-d tree on \emph{Bicycle} (the largest scene across datasets with $\sim$4.2M Voronoi cells and $\sim$63.7M faces), we obtain 1024 partitions with balanced local-cell counts and bounded adjacency.
\cref{fig:bicycle_kdtree_partition_result} plots, per partition, the number of local cells, neighbor cells, adjacency entries, and neighboring clusters.
In the worst case a partition has $N_\text{local}=4102$ local cells, $N_\text{neighbor}=3155$ neighbor cells, and $N_\text{adjacency}=63{,}557$ adjacency entries.
With the compact per-cell layout from ~\cref{subsec:ray_representation},
this corresponds to a per-partition footprint of
$\approx 270$\,kB, leaving roughly 200\,kB space per ray tracer tile for ray payload buffers, and a framebuffer slice.

Octree-based partitioning was also considered. However, it violates either number of partitions or SRAM size limit(See \cref{appendix:scene_partitioning}).
This motivates our choice of k-d tree partitioning plus a separate quadtree routing overlay.

\begin{figure}
\centering
\includegraphics[width=0.99\linewidth, trim={0.5cm 0cm 0.5cm 0cm}, clip]{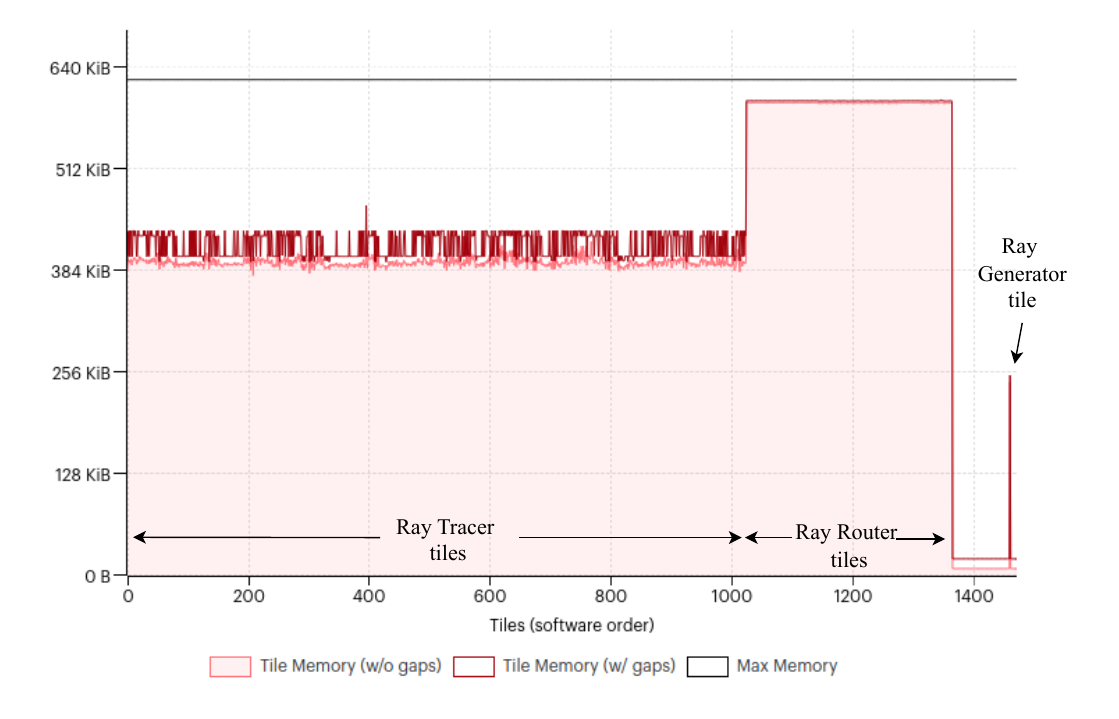}
\caption{PopVision Memory Report. Ray tracers (tiles 0–1023) use up to $\approx 450$\,kB each. Routers (tiles 1024–1364) consume nearly all 624\,kB. The generator (tile 1460) uses $\approx 250$\,kB.}
\label{fig:memory_report}
\end{figure}

PopVision’s Memory Report (\cref{fig:memory_report}) confirms this: router tiles are nearly full ($\approx 624$\,kB), while ray tracers sit around 450\,kB, with sufficient space for ray buffers and the distributed framebuffer slices.

\subsection{Throughput and batching}
\label{subsec:throughput}

We now study end-to-end throughput for interactive rendering.

\paragraph{Device-side loop and repeat count.}
Fusing all subprograms(~\cref{subscn:execution_on_ipu}) into a single device-side loop shifts the majority of runtime from host engine launches and global barriers into tile-local compute-set execution
(Detailed comparison can be found in\cref{appendix:ipu_profiling}).
We further accelerate execution by tuning the \emph{repeat count $RC$}, the number of compute–exchange iterations executed on device before reading back the output image and updating the camera.

\begin{figure}
  \centering
  \includegraphics[width=0.99\linewidth]{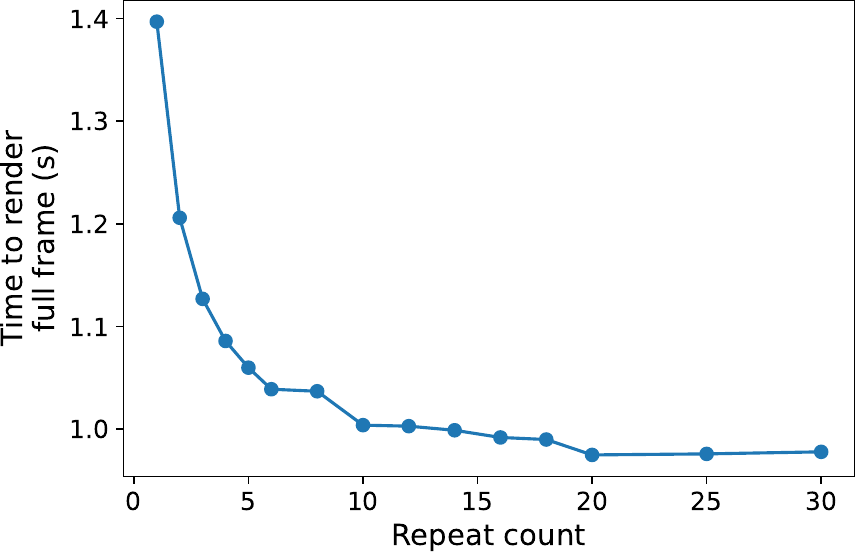}
  \caption{Full-frame rendering time vs repeat count \(RC\) at \(640{\times}480\).
  Speed gain plateaus after $RC{\approx}20$.}
  \label{fig:rc_curve}
\end{figure}

\cref{fig:rc_curve} shows full-frame rendering time versus $RC$ at VGA resolution.
Small $RC$ under-utilizes the IPU due to frequent host interaction; beyond $RC{\approx}20$ returns diminish.
We therefore fix $RC{=}20$ for the remaining experiments.

\paragraph{Row vs column batching.}
The routing overlay cannot receive all rays for a frame at once, so the ray generator emits scan-ordered batches (\cref{subsec:ray_generator}).
We compare column-wise and row-wise batching; detailed sweeps appear in \cref{appendix:batching_sweeps}.

Here, we report the row-wise configuration, which performs slightly better overall.
Table~\ref{tab:rowscan_min} shows results with $R$ rows per batch and a 200-iteration drain after full injection.
For $R{=}1$ we reach $0.952$\,s per frame with $\approx 714$ , IPU iterations per second (IIPS), where one iteration denotes a single compute–exchange cycle. This yields roughly i.e.\ $\approx 1$\,fps at VGA.
Larger batches increase congestion near the top of the tree: the number of iterations required to inject full frame drops, but IPU iterations per second fall and total frame time increases.

\begin{table}[h]
\centering
\caption{Row scanning at \(640{\times}480\) with a 200-iteration drain after injection, rendering at approximately 1 frame per second.}
\label{tab:rowscan_min}
\begin{tabular}{|l|c|c|c|}
\hline
Rows per batch & \(R{=}1\) & \(R{=}2\) & \(R{=}3\) \\
\hline
Rays per batch & 640 & 1280 & 1920 \\
Iterations total & 680 & 500 & 380 \\
Iterations required & 480 & 240 & 160 \\
Time (s) & \textbf{0.952} & 0.986 & 1.027 \\
IIPS & \(\approx 714\) & \(\approx 507\) & \(\approx 370\) \\
\hline
\end{tabular}
\end{table}

For column-wise batching, inserting a small number of idle iterations between batches (\cref{appendix:batching_sweeps,tab:idle_perf_appendix}) helps the routing tree drain and brings performance close to the best row-wise configuration. Full details and drain-tuning sweeps are given in \cref{appendix:batching_sweeps}. Overall, small batches combined with short drain windows offer the best trade-off between throughput and congestion.

\subsection{Routing path length and ray lifetime}
\label{subsec:routing_path_length}

Throughput is ultimately limited not just by how many rays we inject, but by how long each ray takes to finish.
To quantify ray lifetime, we measure routing path length in number of tiles it visits before finishing.
For each rendered view we count, per pixel, how many router hops and tracer hops a ray executes before termination.
These counts jointly characterize
(1) the path length through the quadtree overlay and
(2) the temporal footprint of the ray in the device-side loop, since each hop corresponds to at least one compute–exchange iteration where that ray occupies buffer space.

\begin{figure}
\centering
\begin{subfigure}[t]{0.495\textwidth}
    \centering
    \includegraphics[width=\textwidth]{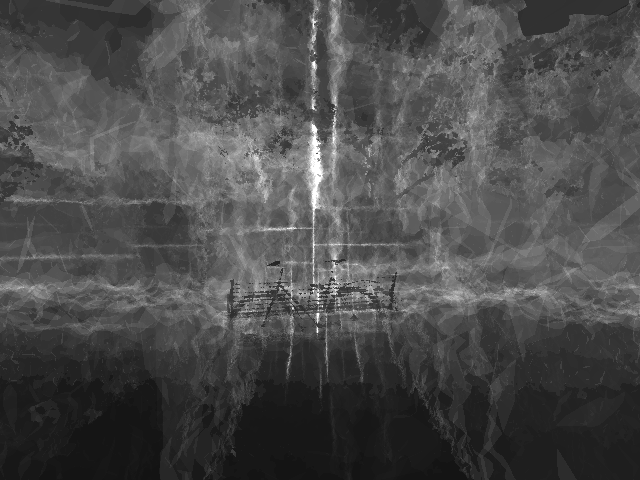}
    \caption{Router hops per pixel (encoded as 0–255; white denotes 255).}
    \label{fig:router_hop_count_bicycle2}
\end{subfigure}\hfill
\begin{subfigure}[t]{0.495\textwidth}
    \centering
    \includegraphics[width=\textwidth]{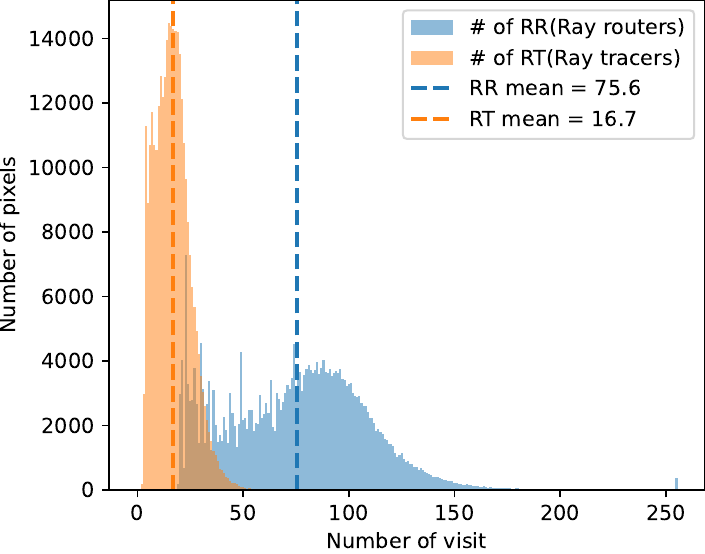}
    \caption{Histogram of router and tracer hops per pixel until termination for the \emph{Bicycle} scene.}
\end{subfigure}
\caption{\emph{Bicycle} scene path statistics. The hop map (a) reveals where rays repeatedly cross partition boundaries, while the histogram (b) summarizes the distribution of router and tracer hops over all pixels, with average values of 75.6 router hops and 16.7 tracer hops per ray.}
\label{fig:hop_count_bicycle2}
\end{figure}

\Cref{fig:hop_count_bicycle2} reports these statistics for a representative \emph{Bicycle} scene.
On average, a ray traverses router tiles 75.6 times and tracer tiles 16.7 times before termination, implying roughly \(75.6 / 16.7 \approx 4.5\) router hops between successive tracer tiles.

The hop-count maps in \cref{fig:router_hop_count_bicycle2} show that most high-hop regions concentrate near geometric boundaries (e.g., foliage silhouettes), where rays repeatedly cross partition interfaces, and along thin bands that likely correspond to congested routes in the quadtree.
Further statistics and results on the \emph{Garden} scene are given in \cref{appendix:hop_statistics}.
Combined with the throughput measurements in \cref{subsec:throughput}, these statistics indicate that routing overhead is modest relative to on-tile traversal, and that most rays terminate after a few tens of tracer hops rather than exploring pathological, deep routes.
In other words, the static overlay keeps routing path length well below the quadtree worst case (10 router hops between tracer tiles in our 5-level tree) and keeps ray lifetime under control.

\subsection{Router capacity and ray payload precision}
\label{subsec:payload_eval}

After program memory allocation, each router tile is left with roughly $\approx 576$\,kB of SRAM to ten buffers (five inputs, five outputs), so each buffer holds about 57.6\,kB of rays.
Router capacity therefore depends directly on the ray payload size from ~\cref{subsec:ray_representation}.
\cref{tab:ray_buffer_capacity_min} reports the implied per-buffer limits for different layouts.

\begin{table}
\centering
\caption{Router buffer capacity vs ray payload size (per buffer \(\approx 57.6~\text{kB} \)).}
\label{tab:ray_buffer_capacity_min}
\begin{tabular}{|l|c|c|}
\hline
Layout (see \cref{subsec:ray_representation}) & Bytes/ray & Max rays\\
\hline
Baseline (origin+dir, fp32) & 48 & \(\sim 1200\) \\
Compact (full precision)    & 28 & \(\sim 2057\) \\
Compact (mixed)             & 24 & \(\sim 2400\) \\
All half (+2\,B pad)        & 20 & \(\sim 2880\) \\
\hline
\end{tabular}
\end{table}

We adopt the 20\,B “all half” layout by default, which nearly doubles in-flight capacity over the naive 48\,B representation.
As shown in~\cref{tab:psnr_ssim_appendix}, this compaction has negligible impact on image quality: PSNR remains above 55\,dB and SSIM $\approx 1.0$ when comparing full-, mixed-, and half-precision payloads, and visual differences are imperceptible.
Together with the throughput results in~\cref{subsec:throughput}, these findings justify aggressively compact ray payloads to trade a small numerical precision loss for significantly higher router throughput.

\subsection{Limitations and artifacts}
\label{subsec:limitations}

Finally, we characterize failure modes that arise from the combination of static routing, tight SRAM budgets, and scan-ordered batching.

\begin{figure}
    \centering
    \includegraphics[width=\linewidth]{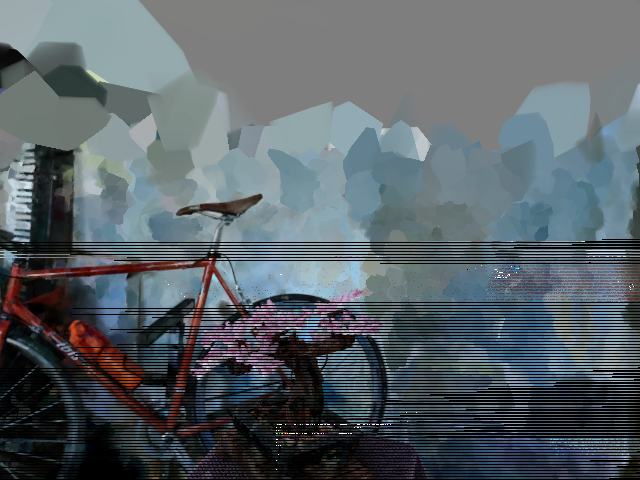}
    \caption{Generator-path saturation in \emph{Bonsai} scene. Overflow appears as missing samples.}
    \label{fig:rg_overflow_artifact_appendix}
\end{figure}

\paragraph{Load imbalance and saturation.}
Because only a subset of scene partitions are visible from any given camera pose, routing and tracing load concentrates along subset of branches of the routing overlay.
\cref{fig:rg_overflow_artifact_appendix} shows a \emph{Bonsai} scene where the generator path saturates: overflow appears as missing samples in the image.
These artifacts motivate using the generator as a controlled spill buffer, together with short drain windows between batches to relieve pressure on congested routes.

\begin{figure}
    \centering
    \includegraphics[width=\linewidth]{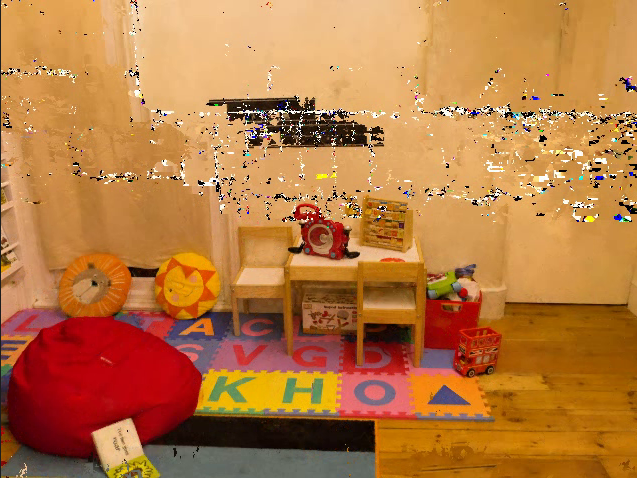} 
    \caption{Ray-origin mismatch artifact in the \emph{Playroom} scene when the camera pose changes mid-batch. Rays launched under the old pose later reconstruct their origins using the updated pose, causing misaligned directions and resulting in black or white speckle.}
    \label{fig:ray_origin_mismatch_artifact_appendix}
\end{figure}

\paragraph{Ray–origin desynchronization.}
To keep payloads small, rays store only pixel coordinates, reconstructing origins from camera parameters broadcast to each tile.
If the camera pose is updated while rays from the previous frame are still being processed in the routing overlay, origins and payloads become inconsistent.
\cref{fig:ray_origin_mismatch_artifact_appendix} shows resulting artifact in the \emph{Playroom} scene when the camera changes mid-batch.
There are two fixes, both with a cost in overall rendering speed.
First, we can defer camera updates until after a global drain, ensuring that no active rays rely on outdated parameters.
Second, we can encode camera-pose information directly in the ray payload, which eliminates desynchronization but increases ray size and reduces effective buffer capacity.

Overall, the evaluation shows that a fully in-SRAM, distributed Radiant Foam renderer on the Mk2 IPU can
(1) closely match the RGB and depth quality of the original Radiant Foam implementation,
(2) fit full Mip-NeRF360 scenes entirely on chip,
and (3) sustain $\approx$1\,fps at VGA resolution under realistic routing and batching constraints, with remaining limitations well characterized and largely addressed by the design choices above.

\section{Conclusion}
We presented an in-SRAM renderer for Radiant Foam on the Graphcore IPU.
By replacing irregular all-to-all messaging with a static quadtree overlay, adopting compact ray payloads, and distributing the output image in screen space across tracer tiles, our system keeps both traversal and accumulation entirely on chip.
At $640\times480$, it achieves interactive throughput ($\approx 1$\,fps) on Mip-NeRF\,360 and Deep Blending scenes while preserving the image and depth quality of the original Radiant Foam implementation.

There are, however, clear trade-offs.
Higher-order spherical harmonics are removed to meet SRAM budgets, and upper-level routers become bottlenecks where traffic aggregates.
Even so, the IPU produces stable, high-quality images and depth at $\approx 1$\,fps on large datasets—a capability, to our knowledge, not previously demonstrated for volumetric rendering on this architecture.
More broadly, as GPUs add fixed-function units they continue to favor uniform SIMD/SIMT execution and triangle-centric pipelines; RT cores do not naturally generalize to volumetric or point-based methods.
Purpose-built accelerators can be very efficient but trade flexibility for long design and verification cycles.
Graph processors such as the IPU offer a different balance: many tiles with local SRAM, explicit message passing, and true MIMD control.
This combination of efficiency and flexibility aligns well with radiance-field workloads that are irregular and data-movement intensive.

\subsection{Future work}
Future work includes relieving congestion near the root of the routing overlay through increasing capacity and congestion-aware scheduling, reducing hop count via alternative overlays and improved partition placement, and repurposing unused tiles to widen bandwidth or stage framebuffer readback.
More efficient storage or lightweight compression could reintroduce higher-order spherical harmonics within SRAM limits to improve image fidelity.
Beyond inference, we currently do not perform training on the IPU; co-designing representation and training procedures for this distributed-SRAM setting is an open direction.
Finally, extending the overlay to multiple IPUs, with inter-IPU routing at the top of the tree, would scale scene size while preserving the in-SRAM execution model.

Overall, this work establishes the feasibility of distributed volumetric rendering on graph processors and provides a foundation for scaling radiance-field scene representations on emerging massively parallel, distributed-memory accelerators.
{
    \small
    \bibliographystyle{ieeenat_fullname}
    \bibliography{main}
}


\clearpage
\setcounter{page}{1}
\appendix
\onecolumn
\begin{center}
\Large
\textbf{\thetitle}\\
\vspace{0.5em}Supplementary Material \\
\end{center}
\vspace{1.0em}

\section{Radiant Foam scene representation}
\label{appendix:radfoam_scene_representation}

\begin{figure}[h]
    \centering
    \begin{subfigure}[t]{0.31\textwidth}
        \centering
        \includegraphics[width=0.995\textwidth]{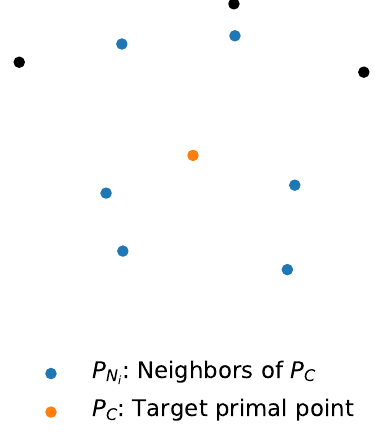}
        \caption{A single Voronoi cell primal point and its neighbors highlighted}
    \end{subfigure}
    \hfill
    \begin{subfigure}[t]{0.31\textwidth}
        \centering
        \includegraphics[width=0.995\textwidth]{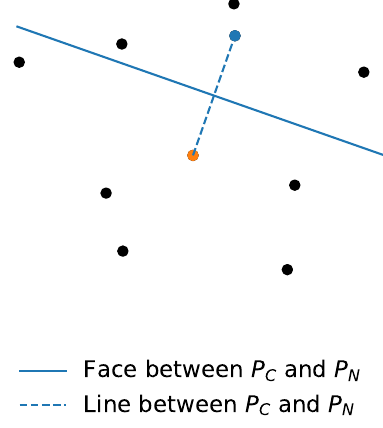}
        \caption{The face between 2 Voronoi cells can be calculated from 2 primal points.}        
        \label{fig:voronoi_reconstruction2}
    \end{subfigure}
    \hfill
    \begin{subfigure}[t]{0.31\textwidth}
        \centering
        \includegraphics[width=0.995\textwidth]{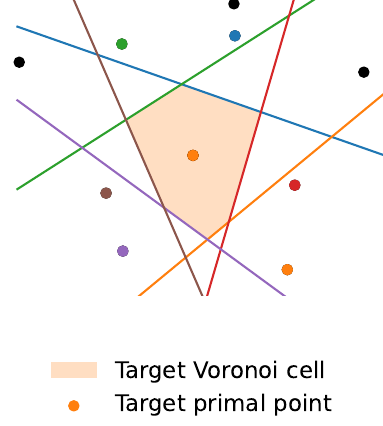}
        \caption{Full Voronoi cell shape is reconstructed by repeating the method for every neighbor.}        
    \end{subfigure}

    \caption{Voronoi cell shape reconstruction from primal points}
    \label{fig:voronoi_reconstruction}
\end{figure}

Radiant Foam represents a scene with three-dimensional Voronoi cells. The Voronoi cell faces do not need to be stored explicitly to explain its shape. The cell shape can be recovered from primal points and its adjacency structure as illustrated in Figure~\ref{fig:voronoi_reconstruction}. 

The face (separating plane) between current and neighboring cells with primal points $p_c$ and $p_n$ is the perpendicular bisector of the segment $[p_c,p_n]$ (Figure~\ref{fig:voronoi_reconstruction2}). Its normal is:
\begin{equation}
n = p_n - p_c,
\end{equation}
and a point on the plane $f_o$ is the midpoint:
\begin{equation}
f_o = \tfrac{1}{2}(p_c + p_n).
\label{eqn:midpoint_equation}
\end{equation}
The plane is then defined by:
\begin{equation}
\Pi = \{\, x \in \mathbb{R}^3 \mid \langle x - f_o, n \rangle = 0 \,\}.
\label{eqn:plane_equation}
\end{equation}
With the convention $n = p_n - p_c$, the half-space belonging to the cell of $p_c$ is:
\begin{equation}
H_c = \{\, x \in \mathbb{R}^3 \mid \langle x - f_o, n \rangle \le 0 \,\}.
\end{equation}
Intersecting these half-spaces over all neighbors reconstructs the full Voronoi cell of $p_c$ (Figure~\ref{fig:voronoi_reconstruction3}).

\clearpage
\section{Scene partitioning: K-d tree vs. Octree}
\label{appendix:scene_partitioning}

We compare KD-tree partitioning (used in the main system) with two octree variants on \emph{Bicycle} from Mip-NeRF 360. KD-tree with 1024 partitions(shards) yields balanced partition sizes and manageable adjacency as shown in the main text. By contrast, octree either explodes the partition count when constraining points per partition, or produces SRAM-exceeding partitions when constraining the number of partitions.

\subsection{Octree with per-partition point cap $\leq$ 5000 points}
\begin{figure}[h]
    \centering
    \includegraphics[width=\linewidth]{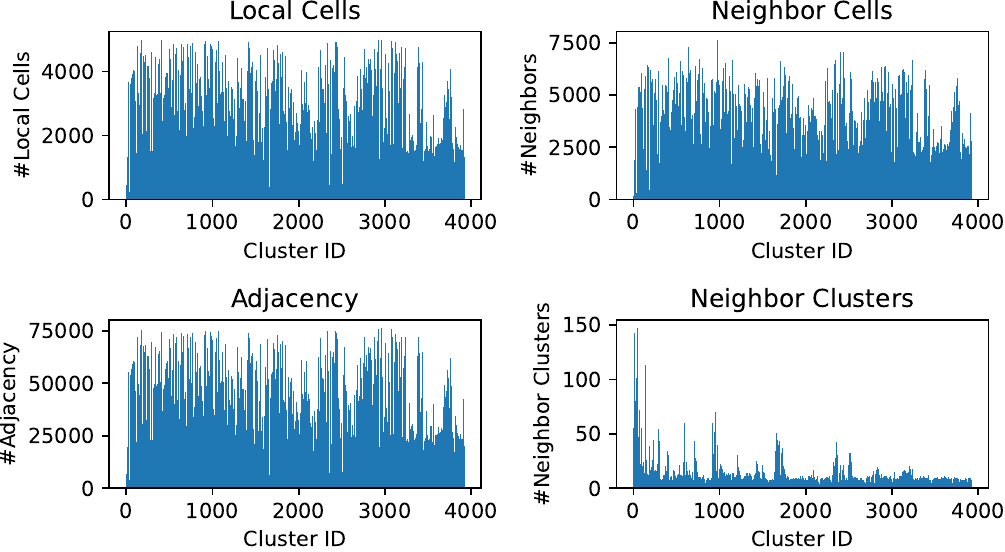}
    \caption{Octree partitioning of \emph{Bicycle} with a maximum of 5000 points per partition. The configuration exceeds the tile count ($\geq$ 3500 partitions).}
    \label{fig:bicycle_octree_maxpts_partition_result_appendix}
\end{figure}

Limiting points per partition greatly increases the number of partitions (\(>\)3500), exceeding the available tiles and making a one-tile-per-partition mapping infeasible as shown in~\cref{fig:bicycle_octree_maxpts_partition_result_appendix}.

\clearpage
\subsection{Octree with Fixed Partition Count (1024 Partitions)}
\begin{figure}[h]
    \centering
    \includegraphics[width=\linewidth]{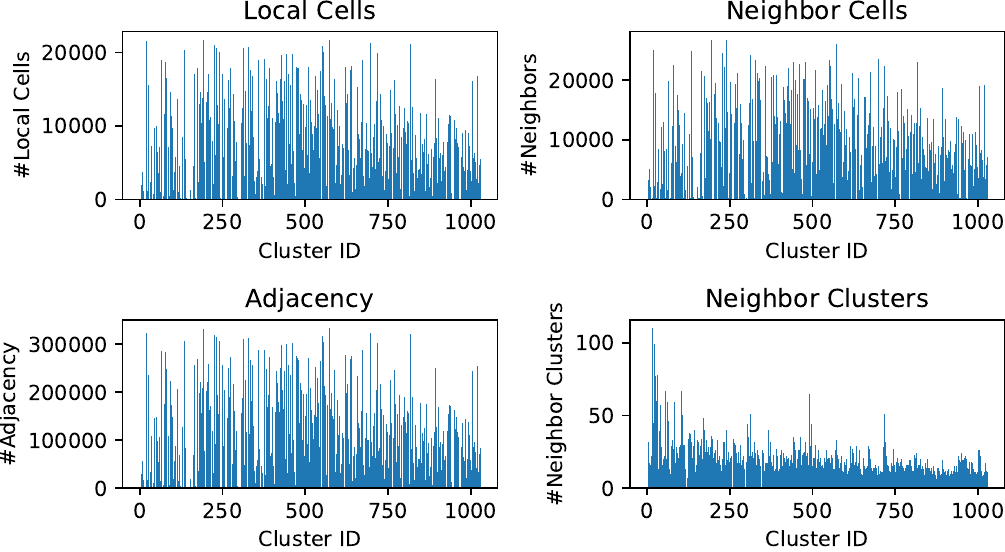}
    \caption{Octree partitioning of \emph{Bicycle} with the number of partitions fixed to 1024. Several partitions exceed on-tile SRAM due to very large local-cell and adjacency counts.}
    \label{fig:bicycle_octree_maxclusters_partition_result_appendix}
\end{figure}

Forcing 1024 partitions yields highly unbalanced partitions: some partitions exceed \(2\times 10^4\) local cells and \(3\times 10^5\) adjacency entries, surpassing 624\,kB SRAM per tile even before queues and workspaces as shown in~\cref{fig:bicycle_octree_maxclusters_partition_result_appendix}.

\clearpage
\section{Execution traces}
\label{appendix:ipu_profiling}
\subsection{Initial execution trace}
\begin{figure}[h]
    \centering
    \includegraphics[width=0.95\linewidth]{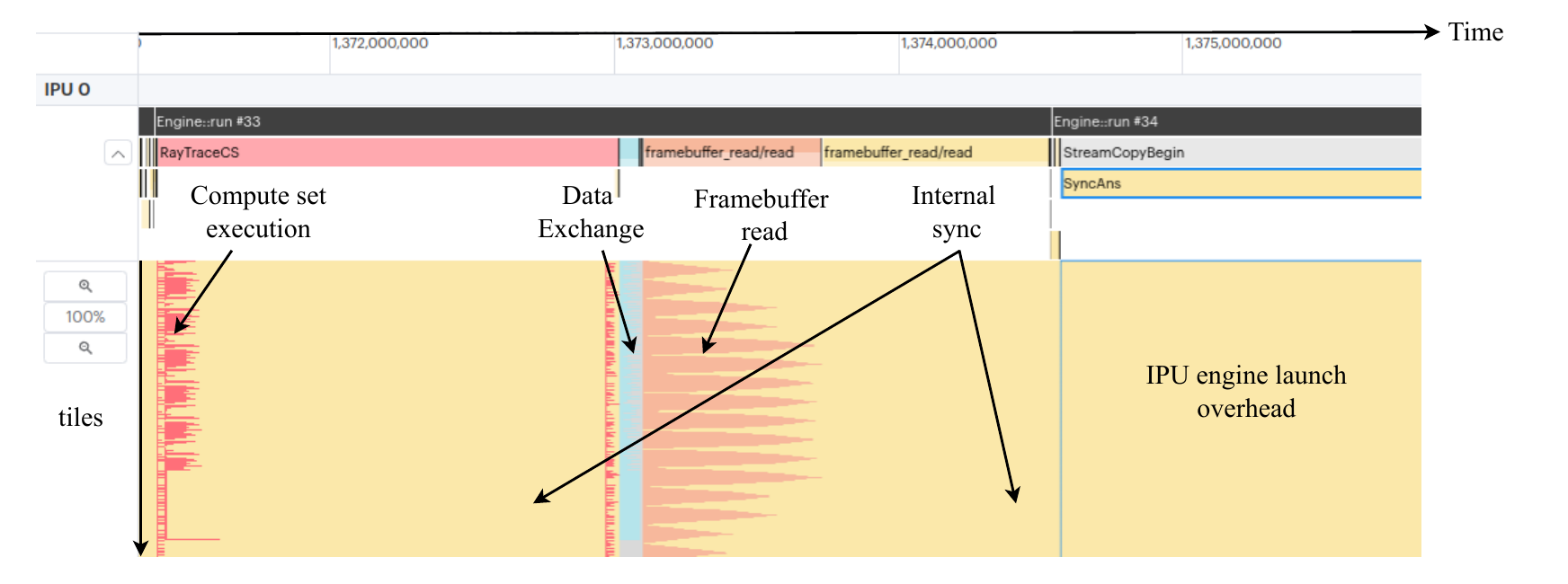}
    \caption{Before fusion: host launches, barriers, and readback dominate.}
    \label{fig:ipu_execution_trace_appendix_before}
\end{figure}

The trace before kernel fusion (Figure~\ref{fig:ipu_execution_trace_appendix_before}) shows that host engine launches, global barriers, and framebuffer readback dominate total time—tile-to-tile exchanges are narrow bands, i.e., relatively cheap. 

\subsection{Optimized execution trace}
\begin{figure}[h]
    \centering
    \includegraphics[width=0.95\linewidth]{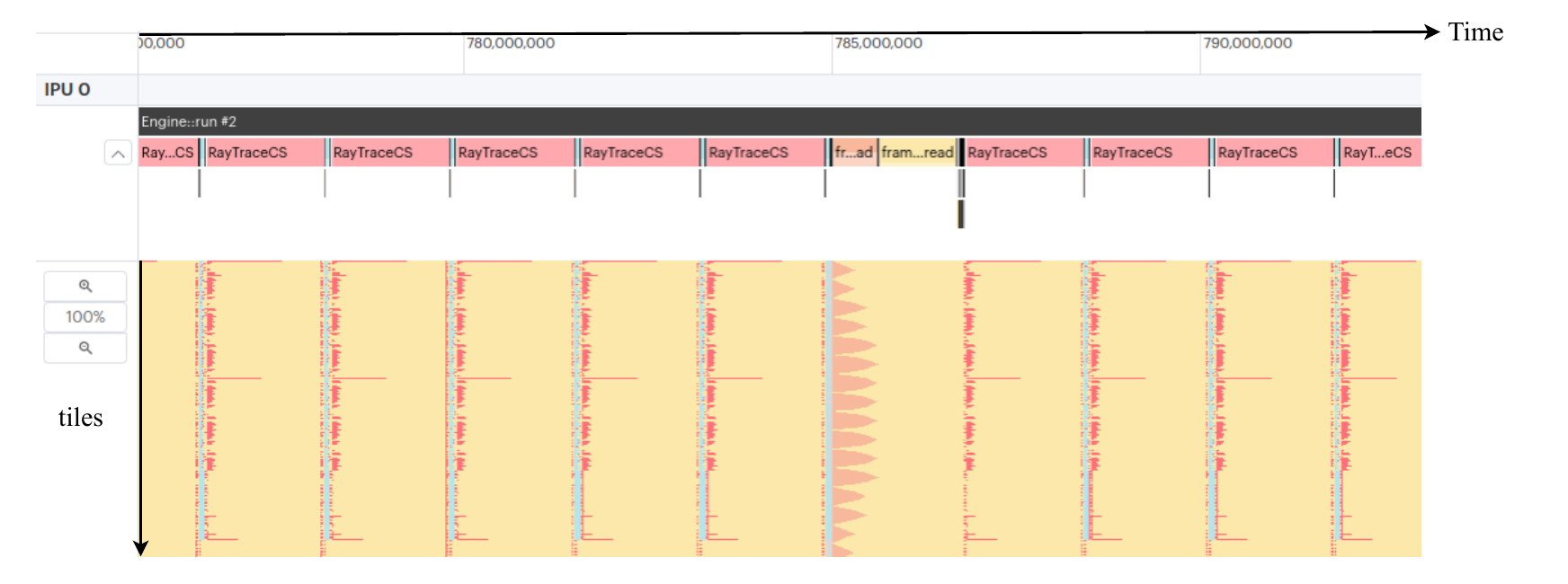}
    \caption{After fusion (device-side loop): runtime becomes compute-set bound.}
    \label{fig:ipu_execution_trace_appendix_after}
\end{figure}

After fusing into a device-side loop (Figure~\ref{fig:ipu_execution_trace_appendix_after}), runtime shifts to compute-set execution, demonstrating that reduced host interaction and batched progress through the routing tree improve utilization and stability.

\clearpage

\section{Routing Path Statistics}
\label{appendix:hop_statistics}

We measure how many times each ray traverses router and tracer tiles before termination. For every pixel in a rendered view, we count the number of router hops and tracer hops along its path and store these as per-pixel maps (values encoded in the range 0–255, with saturation at 255). The resulting hop distributions depend on both scene content and view direction, since different viewpoints activate different regions of the quadtree and the trained model.

\subsection{Bicycle scene}

\begin{figure}[h]
\centering
\begin{subfigure}[t]{0.495\textwidth}
    \centering
    \includegraphics[width=\textwidth]{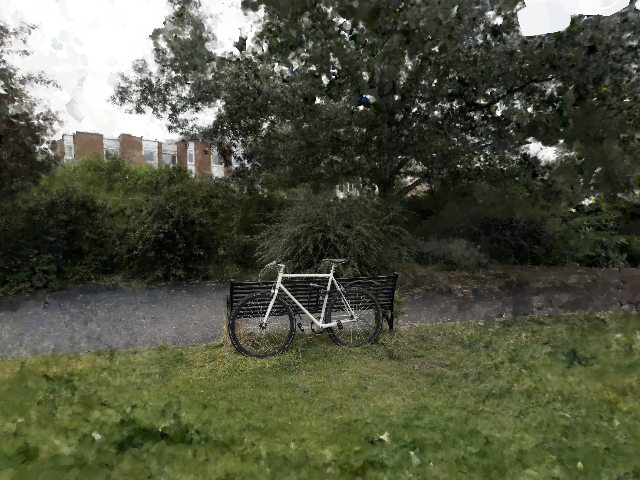}
    \caption{Rendered RGB view of the \emph{Garden} scene.}
\end{subfigure}\hfill
\begin{subfigure}[t]{0.495\textwidth}
    \centering
    \includegraphics[width=\textwidth]{figures/evaluation/hop_count/bicycle_rr_hop_count.png}
    \caption{Router hops per pixel (encoded as 0–255; white denotes 255).}
\end{subfigure}\hfill
\begin{subfigure}[t]{0.495\textwidth}
    \centering
    \includegraphics[width=\textwidth]{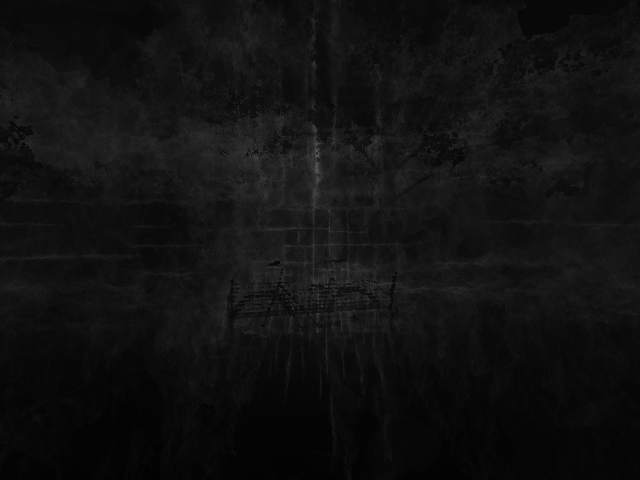}
    \caption{Tracer hops per pixel (encoded as 0–255; white denotes 255).}
\end{subfigure}\hfill
\begin{subfigure}[t]{0.495\textwidth}
    \centering
    \includegraphics[width=\textwidth]{figures/evaluation/hop_count/bicycle_hops_hist.pdf}
    \caption{Histogram of router and tracer hops per pixel until termination.}
\end{subfigure}

\caption{\emph{Bicycle} scene path statistics. Router and tracer hop maps reveal where rays repeatedly cross partition boundaries, while the histogram summarizes the distribution over all pixels.}
\label{fig:hop_count_bicycle}
\end{figure}

Figure~\ref{fig:hop_count_bicycle} shows routing statistics for a representative \emph{Bicycle} view. On average, a pixel’s ray traverses router tiles 75.6 times and tracer tiles 16.7 times before termination, implying roughly \(75.6 / 16.7 \approx 4.5\) router hops between successive tracer tiles. This is well below the quadtree worst case of 10 hops. High-hop regions concentrate near object boundaries (e.g., tree branches and trunk edges), where rays frequently cross partition boundaries.

\clearpage

\subsection{Garden scene}
We do same analysis on \emph{Garden} scene also.

\begin{figure}[h]
\centering
\begin{subfigure}[t]{0.495\textwidth}
    \centering
    \includegraphics[width=\textwidth]{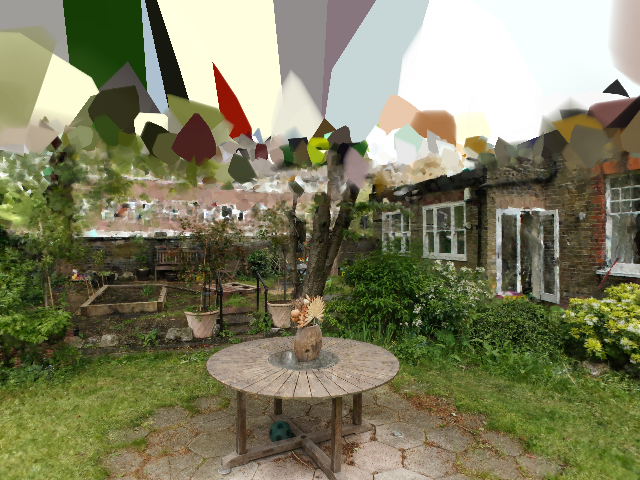}
    \caption{Rendered RGB view.}
\end{subfigure}\hfill
\begin{subfigure}[t]{0.495\textwidth}
    \centering
    \includegraphics[width=\textwidth]{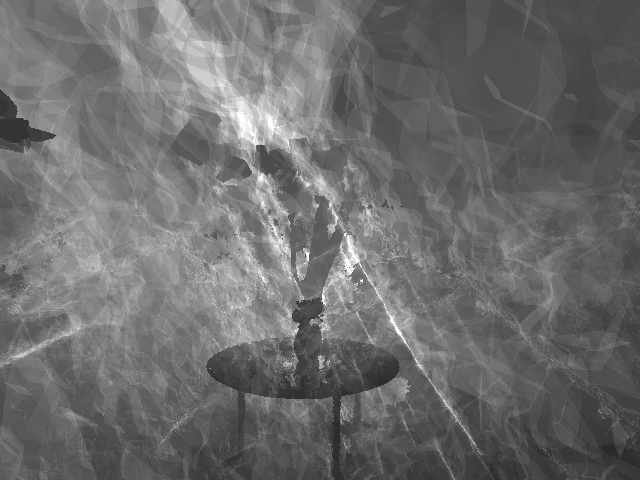}
    \caption{Router hops per pixel (encoded as 0–255; white denotes 255).}
\end{subfigure}\hfill
\begin{subfigure}[t]{0.495\textwidth}
    \centering
    \includegraphics[width=\textwidth]{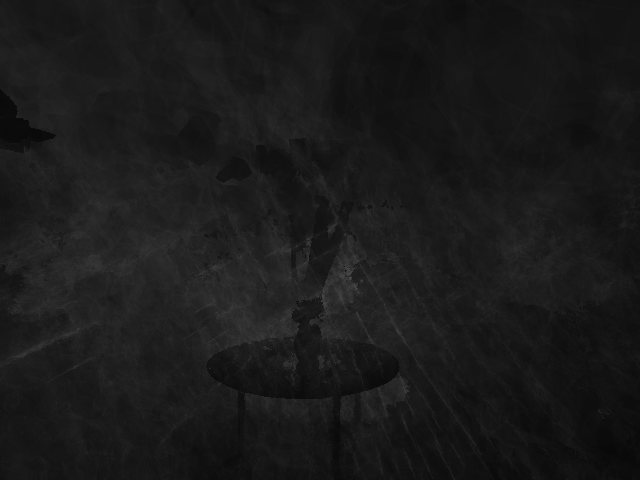}
    \caption{Tracer hops per pixel (encoded as 0–255; white denotes 255).}
\end{subfigure}\hfill
\begin{subfigure}[t]{0.495\textwidth}
    \centering
    \includegraphics[width=\textwidth]{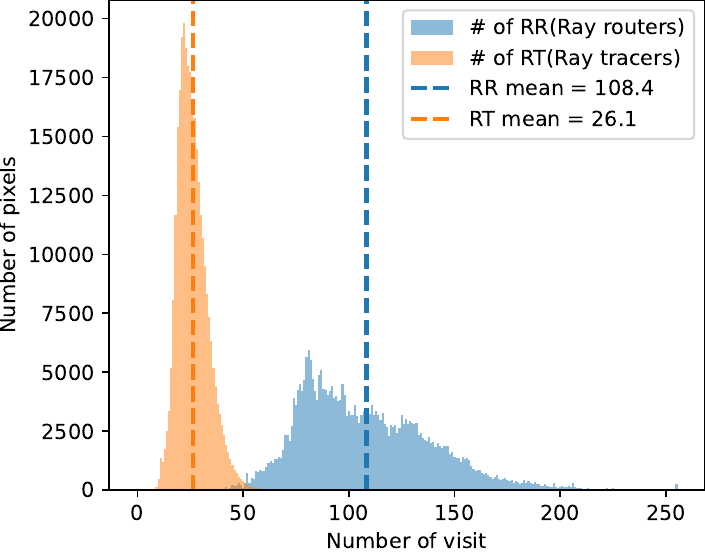}
    \caption{Histogram of router and tracer hops per pixel until termination.}
\end{subfigure}

\caption{\emph{Garden} scene path statistics. Router and tracer hop maps reveal where rays repeatedly cross partition boundaries, while the histogram summarizes the distribution over all pixels.}
\label{fig:hop_count_garden}
\end{figure}

Figure~\ref{fig:hop_count_garden} reports the same statistics for a \emph{Garden} view. On average, a pixel’s ray traverses router tiles 108.4 times and tracer tiles 26.1 times, giving \(108.4 / 26.1 \approx 4.2\) router hops between tracer tiles. Although this scene requires more total hops than \emph{Bicycle}, the average hop distance between tracer tiles remains well below the maximum of 10. We also observe thin high-hop bands in the router map, possibly indicating where rays that passed through congested paths.

\clearpage

\section{Batching sweeps}
\label{appendix:batching_sweeps}

\subsection{Column scanning}
Table~\ref{tab:colscan_perf_appendix} shows column-wise batching result at $640{\times}480$. “Iterations total” far exceeds the “required” minimum as number of columns $C$ grows, revealing congestion near the root (drain time). IPU iterations per second (IIPS) also drops with larger batches, favoring smaller $C$ for higher sustained throughput.

\begin{table}[h]
\centering
\caption{Column scanning at $640{\times}480$ on \emph{Garden}. IIPS: IPU iterations per second.}
\label{tab:colscan_perf_appendix}
\footnotesize
\begin{tabular}{|r|c|c|c|c|}
\hline
Columns/batch $C$ & Rays/batch & Iterations total (required) & Time (s) & IIPS \\
\hline
1 & 480  & 820 (640) & 1.121 & $\approx 731$ \\
2 & 960  & 520 (320) & 1.141 & $\approx 457$ \\
3 & 1440 & 480 (213) & 1.374 & $\approx 349$ \\
4 & 1920 & 460 (160) & 1.433 & $\approx 321$ \\
5 & 2400 & 480 (128) & 2.485 & $\approx 193$ \\
\hline
\end{tabular}
\end{table}

\subsection{Idle iterations between batches}
\begin{figure}[h]
    \centering
    \includegraphics[width=0.85\linewidth]{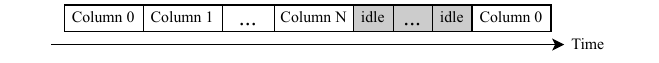}
    \caption{Column scanning with idle iterations inserted between batches.}
    \label{fig:col_scan_idle_appendix}
\end{figure}

Figure~\ref{fig:col_scan_idle_appendix} illustrates the insertion of idle steps between batches, allowing the tree to drain congestion before new rays arrive. Table~\ref{tab:idle_perf_appendix} shows that $+$200 idle iterations strike a good balance: they relieve top-level congestion and reduce frame time, with the best setting at $C{=}2$, $+$200 idle.

\begin{table}[h]
\centering
\caption{One-frame time (s) with idle iterations after each batch. Best at $C{=}2$, $+$200 idle iterations.}
\label{tab:idle_perf_appendix}
\footnotesize
\begin{tabular}{|r|c|c|c|c|c|c|}
\hline
$C$ & Rays & No idle & +100 & +150 & +200 & +225 \\
\hline
1 & 480  & 1.121 & 1.110 & 1.083 & 1.096 & 1.097 \\
2 & 960  & 1.141 & 1.088 & 1.062 & \textbf{0.975} & 0.976 \\
3 & 1440 & 1.374 & 1.278 & 1.163 & 1.083 & 1.007 \\
4 & 1920 & 1.433 & 1.394 & 1.310 & 1.186 & 1.129 \\
5 & 2400 & 2.485 & 1.691 & 1.637 & 1.410 & 1.382 \\
\hline
\end{tabular}
\end{table}

\clearpage

\section{Additional qualitative results}
\label{appendix:additional_qualitative}

\begin{figure}[h]
    \centering
    \begin{subfigure}[t]{0.449\textwidth}
        \centering
        \includegraphics[width=\textwidth]{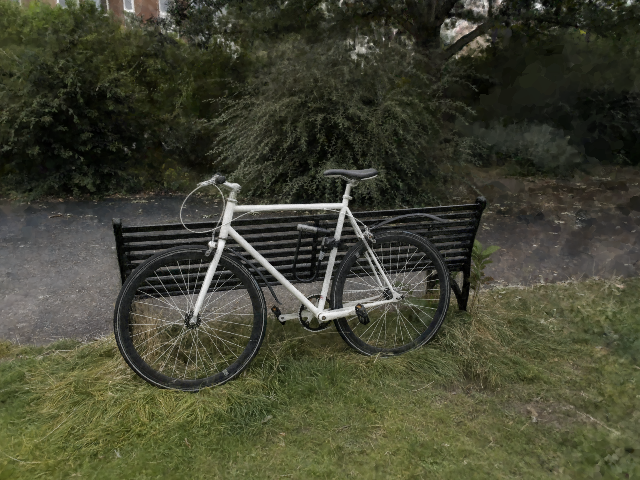}
        \caption{Bicycle (RGB)}
    \end{subfigure}\hfill
    \begin{subfigure}[t]{0.449\textwidth}
        \centering
        \includegraphics[width=\textwidth]{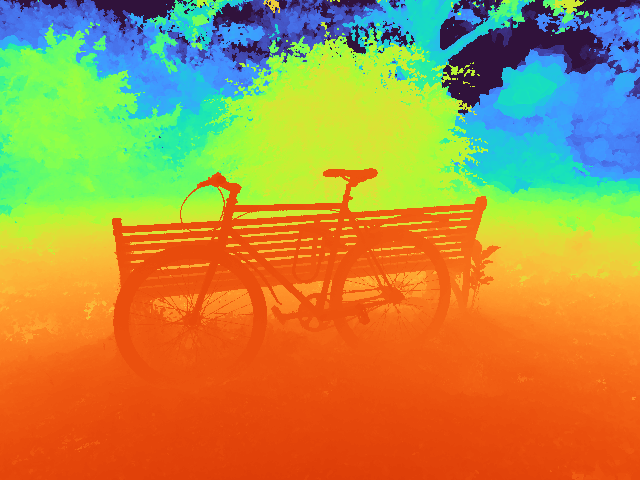}
        \caption{Bicycle (Depth)}
    \end{subfigure}

    \vspace{0.35em}

    \begin{subfigure}[t]{0.449\textwidth}
        \centering
        \includegraphics[width=\textwidth]{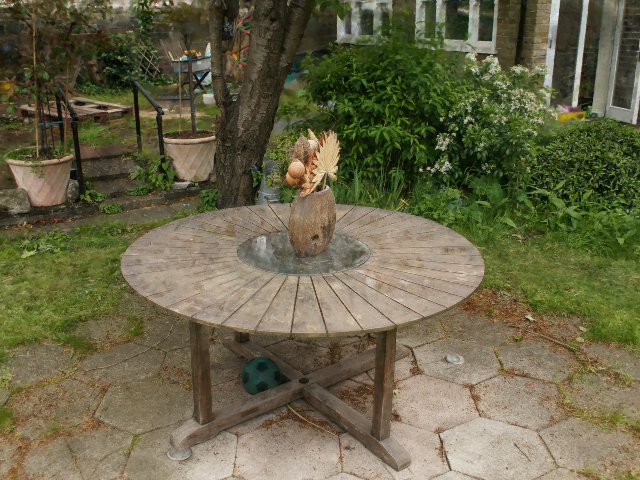}
        \caption{Garden (RGB)}
    \end{subfigure}\hfill
    \begin{subfigure}[t]{0.449\textwidth}
        \centering
        \includegraphics[width=\textwidth]{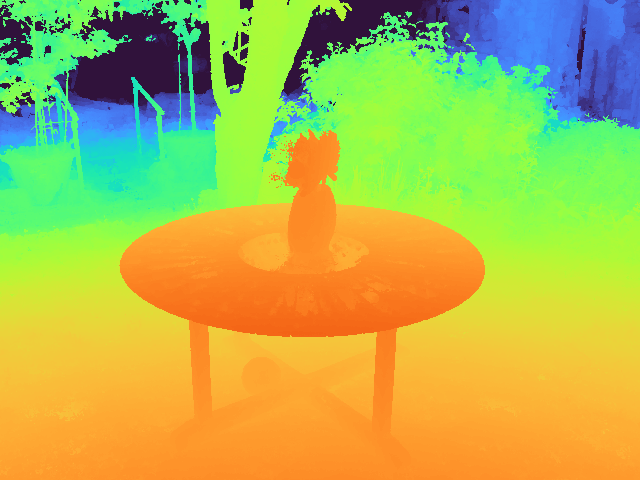}
        \caption{Garden (Depth)}
    \end{subfigure}

    \vspace{0.35em}

    \begin{subfigure}[t]{0.449\textwidth}
        \centering
        \includegraphics[width=\textwidth]{figures/evaluation/stump_rgb_image.png}
        \caption{Stump (RGB)}
    \end{subfigure}\hfill
    \begin{subfigure}[t]{0.449\textwidth}
        \centering
        \includegraphics[width=\textwidth]{figures/evaluation/stump_depth_image.png}
        \caption{Stump (Depth)}
    \end{subfigure}
    \caption{RGB renderings (left) and depth maps (right) at 640$\times$480 resolution for the \emph{Bicycle}, \emph{Garden}, and \emph{Stump} scenes. Depth is computed at the 50\% transmittance quantile along each ray.}
    \label{fig:rendering_result_depth_appendix}
\end{figure}

\begin{figure}[h]
    \centering
    \begin{subfigure}[t]{0.449\textwidth}
        \centering
        \includegraphics[width=\textwidth]{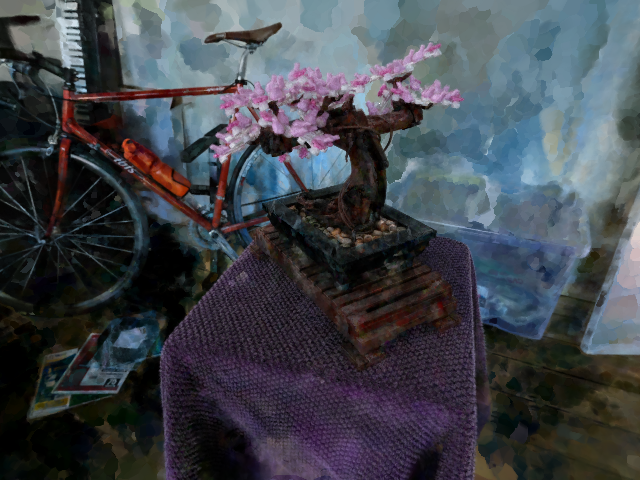}
        \caption{Bonsai (RGB)}
    \end{subfigure}\hfill
    \begin{subfigure}[t]{0.449\textwidth}
        \centering
        \includegraphics[width=\textwidth]{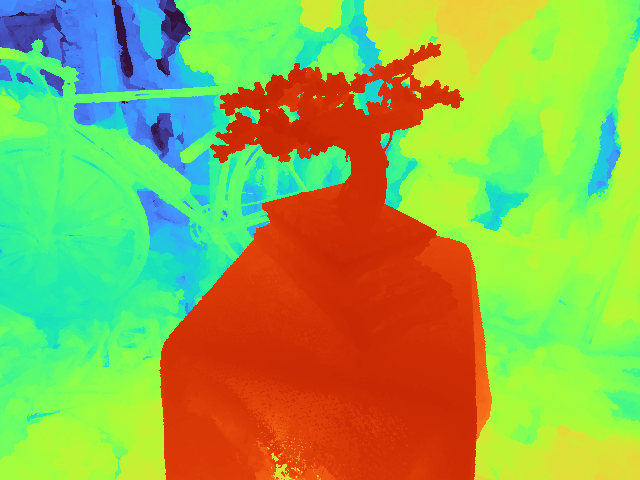}
        \caption{Bonsai (Depth)}
    \end{subfigure}

    \vspace{0.35em}

    \begin{subfigure}[t]{0.449\textwidth}
        \centering
        \includegraphics[width=\textwidth]{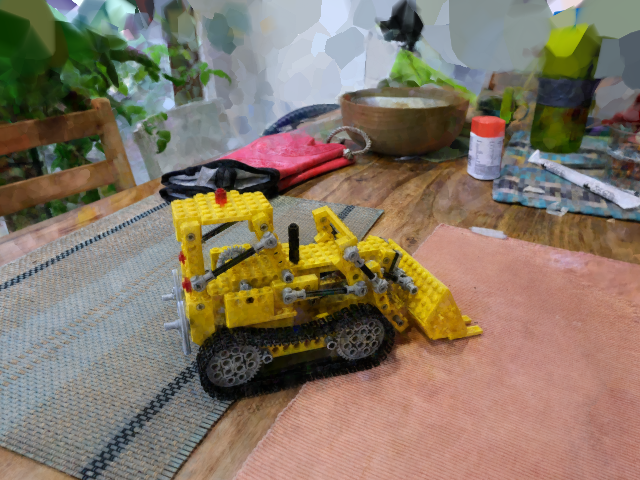}
        \caption{Kitchen (RGB)}
    \end{subfigure}\hfill
    \begin{subfigure}[t]{0.449\textwidth}
        \centering
        \includegraphics[width=\textwidth]{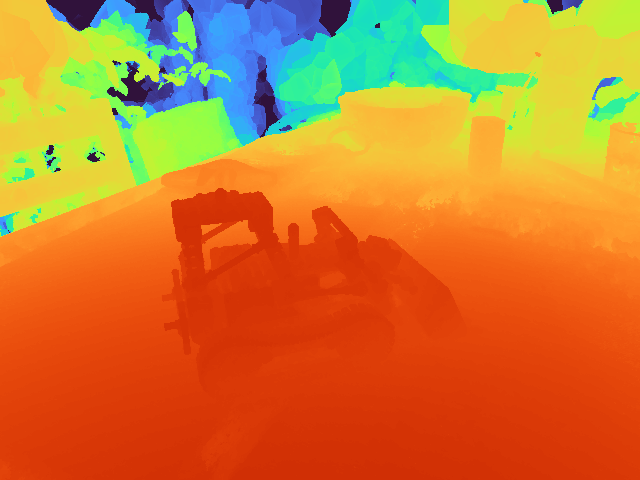}
        \caption{Kitchen (Depth)}
    \end{subfigure}

    \vspace{0.35em}

    \begin{subfigure}[t]{0.449\textwidth}
        \centering
        \includegraphics[width=\textwidth]{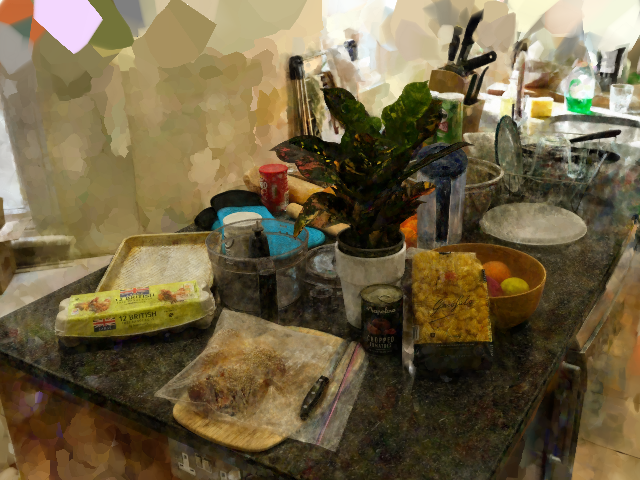}
        \caption{Counter (RGB)}
    \end{subfigure}\hfill
    \begin{subfigure}[t]{0.449\textwidth}
        \centering
        \includegraphics[width=\textwidth]{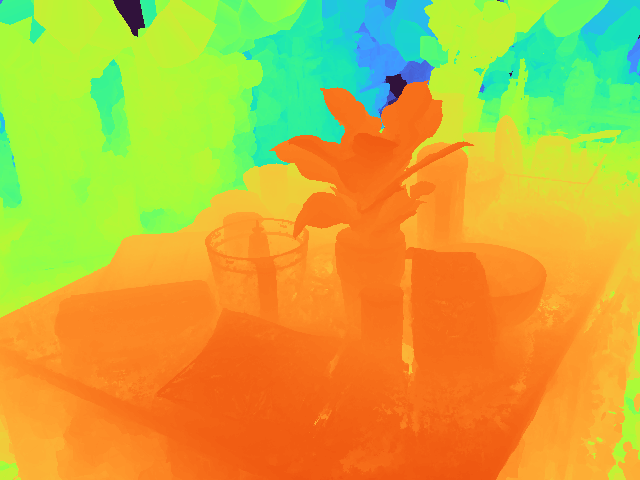}
        \caption{Counter (Depth)}
    \end{subfigure}
    \caption{RGB renderings (left) and depth maps (right) at 640$\times$480 resolution for the \emph{Bonsai}, \emph{Kitchen}, and \emph{Counter} scenes. Depth is computed at the 50\% transmittance quantile.}

    \label{fig:rendering_result_depth_appendix2}
\end{figure}

Figures~\ref{fig:rendering_result_depth_appendix} and~\ref{fig:rendering_result_depth_appendix2} show paired RGB renderings and depth maps for six scenes. Depth is computed at the 50\% transmittance quantile, producing smooth, geometry-aligned results.

\end{document}